\documentclass[]{fairmeta}
\usepackage[]{todonotes}
\usepackage[sort&compress,numbers,super]{natbib}
\usepackage{url}
\usepackage{chngcntr}

\title{Open Materials 2024 (OMat24) Inorganic Materials Dataset and Models}

\author[]{Luis Barroso-Luque}
\author[]{Muhammed Shuaibi}
\author[]{Xiang Fu}
\author[]{Brandon M. Wood}
\author[]{Misko Dzamba}
\author[]{Meng Gao}
\author[]{Ammar Rizvi}
\author[]{Matt Uyttendaele}
\author[]{C.~Lawrence Zitnick}
\author[]{Zachary W. Ulissi}

\affiliation{Fundamental AI Research (FAIR) at Meta}

\abstract{\noindent  The discovery and simulation of inorganic materials is core to diverse applications from climate change to semiconductor manufacturing. Artificial intelligence (AI) has the potential to dramatically accelerate materials simulation, discovery and design. Although considerable progress has been made in developing training datasets and machine learning interatomic potential (MLIP) architectures, the state-of-the-art in openly available and reproducible datasets and models lagged behind proprietary models. To address this issue, we present the Open Materials 2024 (OMat24) dataset, comprising over 110M DFT calculations across diverse chemistries, materials, and configurations. MLIP models trained on OMat24 achieve leading performance on the Matbench Discovery leaderboard, surpassing previous models with F1 scores above 0.9 for stability and 20 meV/atom accuracy for formation energy. Models trained on OMat24 also exhibit the highest accuracy in newly developed thermal conductivity and phonon prediction task benchmarks. We show that OMat24's diversity corrects the consistent softening bias of prior models trained on less diverse datasets, which systematically underpredicted energy, forces, and derivative properties like phonons. The OMat24 dataset has allowed the research community to develop better model architectures, resulting in a step-change improvement inorganic material property prediction accuracy.
}

\date{\today}

\correspondence{Luis Barroso-Luque (\email{lbluque@meta.com}), Z. Ulissi (\email{zulissi@meta.com}) }

\metadata[Code]{\url{https://github.com/FAIR-Chem/fairchem}, MIT license}
\metadata[Dataset]{\url{https://huggingface.co/datasets/fairchem/OMAT24}, Creative Commons 4.0 License}
\metadata[Checkpoints]{\url{https://huggingface.co/fairchem/OMAT24}, Custom License}
\metadata[Blogpost]{\url{https://ai.meta.com/blog/fair-news-segment-anything-2-1-meta-spirit-lm-layer-skip-salsa-sona}}

\begin{document}

\maketitle

\section{Introduction}

The discovery of new materials lies at the foundation of many pressing global problems. This includes finding new materials for renewable energy storage, producing carbon neutral fuels \cite{oc20,zitnick2020introduction,singh2019robust}, or the direct capture of CO$_2$ from the air\cite{sriram_open_2024}, among many others \cite{choudhary2022recent, ceder_1998, biomat_2015, carbtree_comp_mat, pogue_2023}. The search space for new materials is enormous, making the ability to efficiently screen new materials essential. This is challenging for both traditional computational methods and experimental material science methods. Machine Learning (ML) models offer the potential to dramatically improve the efficiency of computational materials simulations compared to traditional methods such as Density Functional Theory (DFT), and address many scientific questions that have been difficult to answer systematically due to their high computational cost.

Recently, there has been a surge in the development of Machine Learning Interatomic Potentials (MLIPs)---surrogates for Density Functional Theory (DFT)---with improving ability to generalize across materials \cite{merchant_scaling_2023, batatia2023foundation, Batatia2022mace, nequip_2022, chen_universal_2022, choudhary_alignn, deng_chgnet_2023}. As in other domains, training large and generalizable models is limited by the availability of large and diverse datasets \cite{krizhevsky2012imagenet, kaplan2020scaling}. The computational materials community has benefited from open \textit{ab-initio} databases such as the Materials Project \cite{jain2013commentary}, Alexandria \cite{schmidt_machine_2023}, OQMD \cite{saal_materials_2013}, and AFlow \cite{curtarolo_aflow_2012}, which contain near-stable structures across diverse compositions and prototypes. While models trained on these datasets have grown in accuracy and size, they frequently extrapolate poorly to configurations that deviate substantially from the near-equilibrium training data, limiting their usefulness for simulations under relevant conditions---likeley due to insufficiently diverse training data \cite{dengSystematicSofteningUniversal2025}. Recent proprietary models and datasets have claimed to address some of these issues \cite{yang2024mattersim, merchant_scaling_2023}, but without access to the underlying training data, open science efforts cannot benefit from or independently replicate these results.

We introduce the Open Materials 2024 (OMat24) dataset (Figure \ref{fig:omat-overview}) to improve the generalization of MLIPs across a wide range of materials and to reduce systematic softening. OMat24 contains over 100 million single-point DFT calculations sampled from diverse non-equilibrium atomic configurations and elemental compositions for inorganic bulk materials, building upon public datasets such as MPtrj \cite{deng_chgnet_2023}, the Materials Project \cite{Jain2013}, and Alexandria \cite{schmidt_machine_2023}, which contain equilibrium or near-equilibrium configurations \cite{jain2013commentary}. We pretrain several variants of the EquiformerV2 \cite{liao2024equiformerv2, liao2024generalizing} and eSEN \cite{fu2025learning} architectures on OMat24 and show that, after fine-tuning to account for differences in DFT settings, state-of-the-art results are achieved on the popular MatBench Discovery leaderboard \cite{riebesellFrameworkEvaluateMachine2025}. Since its release, all top models \cite{rhodesOrbv3AtomisticSimulation2025, park2024scalable, bochkarevGraphAtomicCluster2024, wang2025graph} on the leaderboard have adopted the dataset. We show that models trained or pretrained on OMat24 display reduced or no systematic softening, and that these improvements are consistent across different model architectures.
\section{Results}

\subsection{The OMat24 dataset}

The OMat24 dataset builds upon valuable insights from previous team initiatives such as the Open Catalyst Project\cite{oc20} and OpenDAC\cite{sriram_open_2024}, community contributions like MPTrj\cite{deng_chgnet_2023} and Alexandria\cite{schmidt_machine_2023}, as well as proprietary releases. Leveraging these learnings, OMat24 was specifically designed to advance MLIP training across inorganic materials and overcome challenges faced by earlier computational materials datasets.

\begin{itemize}
    \item Scale: OMat24 is a large-scale public ab-initio training dataset for inorganic materials up to 2 orders of magnitude larger than other openly available datasets in terms of number of single-point calculations.
    \item Diversity: OMat24 has compositional diversity across the periodic table thanks to community efforts to prototype and catalog hypothetical materials, and contains a large fraction of far-from-equilibrium structures.
    \item Open-science: OMat24 is released under a permissive CC-by license for the benefit of the entire inorganic materials community to build upon. 
\end{itemize}

\begin{figure*}[h!]
    \centering
    \includegraphics[width=0.85\textwidth]{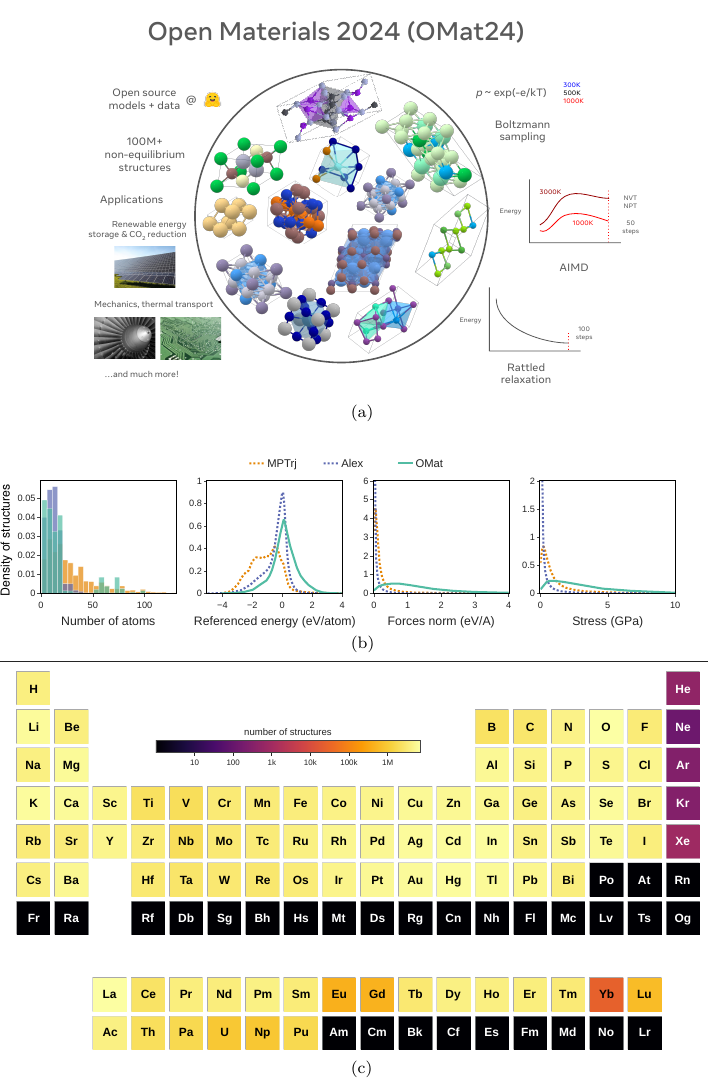}
    \caption{\textbf{OMat24 dataset overview.} (a) Dataset schematic including generation, application areas, and sampling strategies: rattled Boltzmann, rattled relax and AIMD. Structures shown are a random sample across the different sampling strategies. (b) Distribution of frequency density of structures in terms of the number of atoms per structure, the energy per atom referenced with respect to elementary bulk phases, forces norm and max absolute stress element for structures in the MPtrj, Alexandria and OMat24 datasets. (c) Distribution of elements present in the OMat24 dataset}
    \label{fig:omat-overview}
\end{figure*}

The OMat24 dataset includes a total of 118 million structures labeled with energy, forces and cell stress. The number of atoms per structure ranges from 1 to 100 atoms per structure. A histogram of the number of atoms per structure is shown in Figure \ref{fig:omat-overview}b. The majority of OMat24 structures have less than 20 atoms per structure as a direct result of starting from structures in the Alexandria dataset, which are predominantly structures with 16 or less atoms per structure. The OMat24 structures with more than 50 atoms per structure are the larger input structures that were used for AIMD. Details of the sampling strategy and DFT calculations are included in the Methods section. 

OMat24 was designed to train MLIP models that predict both equilibrium and non-equilibrium properties. It includes equilibrium and non-equilibrium structures, yielding broader label distributions than relaxation-only datasets. We confirm this by comparing final label distributions to MPtrj and Alexandria, shown in Figure \ref{fig:omat-overview}b, plotting energies relative to elemental bulk phases computed with each dataset’s DFT settings. OMat24 shows slightly higher energies than Alexandria (the source of input structures) and substantially higher energies than MPtrj, along with notably wider force and maximum cell-stress distributions than both datasets.

OMat24’s elemental coverage spans most of the periodic table (Figure \ref{fig:omat-overview}c) and closely matches Alexandria and MPtrj (Supplementary Figures \ref{fig:alex-elements}, \ref{fig:mptrj-elements}), with oxides over-represented due to their prevalence in open data. Finally, OMat24 exhibits greater diversity of local atomic environments, reflected in higher and more uniform element-pair counts within 3.5 Å (Supplementary Figure \ref{fig:pair-counts}), exceeding both MPtrj and Alexandria.

\subsubsection{OMat24 train, validation, and test splits\label{sec:dataset_split}}

OMat24 is divided into explicit training, validation and test splits to ensure consistent training and evaluation by the community. Training and validation splits are released to allow for model development and iteration. The test set is divided into four different splits. The first split (WBM Test) is to ensure that the training dataset does not overlap with the Matbench Discovery leaderboard \cite{riebesellFrameworkEvaluateMachine2025} created from the WBM dataset \cite{wang_predicting_2021}. The other three splits measure the accuracy of the models on in-domain training data (ID) and the ability of the models to generalize to out-of-distribution compositions (OOD-Composition) and elemental compositions (OOD-Element). 

The WBM Test split was created using the AFLOW structure prototype labels \cite{mehl_aflow_2017} in the \textsc{aviary} package \cite{goodall2022rapid}. The prototype label of a structure is a standardized way to classify crystal structures by elemental stoichiometry, space group, Pearson symbol, and Wyckoff positions \cite{mehl_aflow_2017}.  The split includes all OMat24 structures that were generated starting from an Alexandria relaxed structure with a prototype label matching any of prototype labels from the initial or relaxed structures included in the WBM dataset. Additionally, all OMat24 structures with a prototype label matching an initial or relaxed WBM structure are also included in the WBM Test split. The WBM Test split includes a total of 5.3 million structures. Note, filtering the dataset and creating this test split is important to ensure that there was no inadvertant data leakage from the training data to the final Matbench Discovery results, as there is overlap in materials between the Alexandria and WBM datasets.

The OOD-Composition split is constructed by selecting approximately $\sim$5,000 unique elemental compositions and assigning all structures with matching compositions to the split (573,000 structures). The OOD-Elemental split similarly selects $\sim$3,000 unique element combinations and includes all matching structures (619,000 structures). A limitation of these splits is the possibility of naive OOD cases (e.g., dilute substitutions or elemental doping) where small stoichiometric changes do not meaningfully increase chemical complexity, though the dataset’s relatively small system sizes and lack of dopants limit such examples. In contrast, the WBM test split—separated by unique anonymized formulas and space groups—consistently yields the highest prediction errors (Supplementary Table \ref{tab:omat-test}), suggesting it represents the most important distribution shift and a more informative test of model generalization.

The training, validation and ID test splits includes all remaining OMat24 structures after creating the 3 test splits described above, and includes a total of 111 million structures. We randomly split this dataset into a training, validation and ID test split containing 100 million, 5 million and 5 million structures respectively for the model training in this work. All dataset split sizes are listed in Table \ref{tab:splits}.

\clearpage

\begin{table*}[hb!]
\centering
\caption{\textbf{OMat24 dataset splits.} Size of the OMat24 train, validation and test dataset splits.\label{tab:splits}}
\begin{tabular}{l|c|c}\toprule
Split &Size &Fraction \% \\\midrule
Train &100,824,585 &85.3 \\
Validation &5,320,549 &4.5 \\
WBM Test &5,373,339 &4.5 \\
ID Test &5,453,320 &4.6 \\
OOD Composition Test &573,301 &0.5 \\
OOD Element Test &619,021 &0.5 \\
\bottomrule
\end{tabular}
\end{table*}

We have also released a development subset of the full dataset, referred to as OMat24-1M, which comprises approximately one million training structures randomly selected from the complete training split. The corresponding validation and test sets contain a number of structures chosen to maintain the same train-to-validation and train-to-test ratios as the full dataset. The OMat24-1M development dataset is particularly useful for rapid training iterations, ablation studies, and hyperparameter optimization.

\subsubsection{Dataset limitations \label{sec:dataset-limitations}}
The OMat24 dataset is  is one of the largest open datasets of its kind for training DFT surrogate models for materials. However, the dataset has limitations similar to many high-throughput datasets that impact the predictions of models trained using the dataset. OMat24 is calculated with PBE and PBE+U levels of DFT, which includes inherent errors in their approximation and resulting calculations \cite{perdew1996generalized} that are addressed to some extent in other functionals such as PBEsol \cite{perdew_restoring_2008}, SCAN \cite{sun_strongly_2015}, r2SCAN \cite{furness_accurate_2020} or hybrid functionals. The OMat24 dataset includes only periodic bulk structures and excludes important effects from point defects, surfaces, non-stoichiometry, and lower dimensional structures. Finally, the OMat24 dataset includes a small fraction of structural relaxations (45,000 total relaxations) starting from distorted relaxed structures in the Alexandria dataset, and does not provide additional or novel information about stable structures.

\begin{figure*}
    \centering
    \includegraphics[width=\textwidth]{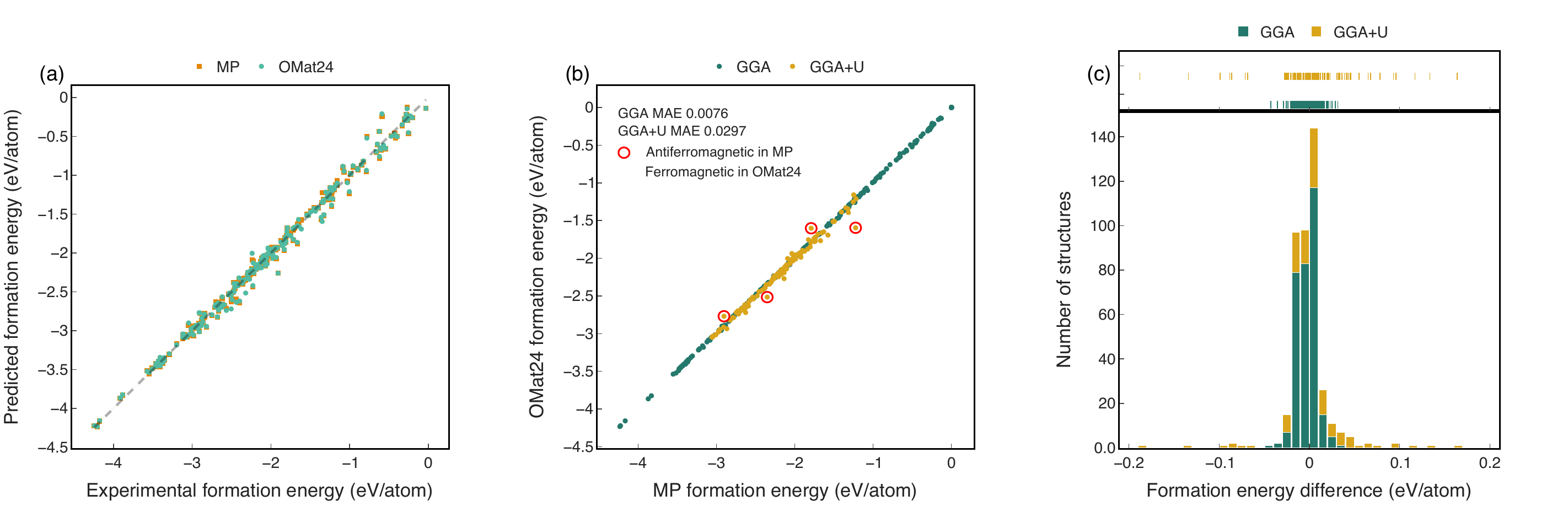}
    \caption{\textbf{OMat24 and Materials Project DFT formation energy parity.} Formation energy computed with Materials Project PBE settings and OMat24 DFT settings for 423 selected compounds from Materials Project. (a) Parity plot of predicted and experimental formation energy. (b) Parity plot between predicted formation energies using OMat24 and MP DFT settings.} (c) Histogram of energy differences. Formation energies are calculated using MP2020 GGA/GGA+U mixing corrections and anion composition corrections. MP corrections are taken directly from MP. OMat24 corrections are fitted following the same procedure using DFT data calculated with OMat24 settings.
    \label{fig:mp-omat-energy}
\end{figure*}

The calculations in OMat24 differ from those found in the Materials Project PBE and PBE+U calculations. Care must be taken when mixing calculations for analysis or training models. Although the difference in settings is small (pseudopotentials in version 5.4 and the choice of pseudopotential for Yb and W), predictions of total and formation energies can differ. To illustrate these subtleties, we compare calculated formation energies for MP and OMat24 DFT settings in Figure \ref{fig:mp-omat-energy} using a set 423 compounds used to fit the MP2020 corrections \cite{wang_framework_2021}. We use the original MP calculations and compare to calculated energies with OMat24 DFT settings. In order to compute formation energies, we computed elemental references and fit anion and GGA/GGA+U corrections following the methodology used in the \textsc{Materials2020Compatibility} class in \textsc{pymatgen}\cite{wang_framework_2021}. Additional details about the calculations and mixing corrections are given in Supplementary Section \ref{sec:omat-refs}.

Figure \ref{fig:mp-omat-energy} shows parity plots for DFT-predicted formation energies of the 423 compounds. Figure \ref{fig:mp-omat-energy}b depicts the distribution of differences between MP and OMat24 DFT formation energy predictions. Most differences are under 20 meV/atom and the total mean absolute difference is 13.5 meV/atom, consistent with expected changes from updated pseudopotentials. Deviations above 50 meV/atom mainly arise in GGA+U calculations for magnetic compounds. These differences primarily arise from variations in magnetic ordering: OMat24 initializes all structures as ferromagnetic, whereas some Materials Project (MP) calculations start with lower-energy magnetic states. Consequently, the final magnetic states may differ between the two. The largest differences, marked with red circles in Figure \ref{fig:mp-omat-energy}b, correspond to compounds that are antiferromagnetic in MP but ferromagnetic in OMat24.

A key limitation in using available DFT datasets to train MLIPs is the aforementioned fixed initialization of magnetic ordering, which does not guarantee that the resulting calculation corresponds to the true magnetic ground state. For instance, models trained on the OMat24 dataset typically predict formation energies for specific magnetic configurations---usually ferromagnetic---rather than the actual magnetic ground states. Accurately computing formation energies of magnetic materials with MLIPs requires careful consideration of the energy differences among various magnetic orderings. Accurate computation of formation energies for magnetic materials using MLIPs requires explicit consideration of the energy differences between possible magnetic orderings. Addressing magnetic structure in MLIPs remains an active area of research, involving both the generation of balanced and informative training datasets\cite{sanspeurCircumventingDataImbalance2024} and advancements in model architecture and training strategies\cite{deng_chgnet_2023, xuSpininformedUniversalGraph2025}.

Finally, the OMat24 dataset was generated using convergence criteria aligned with Materials Project PBE calculations \cite{jain2013commentary}, striking a balance between computational efficiency and accuracy to support large-scale data generation within available compute resources. Later analyses has identified AIMD subsplits exhibiting nonzero drift forces, suggesting that tighter convergence could be beneficial \cite{kuryla2025accuratedftforces}. However, as shown in the Supplementary Table \ref{tab:force-convergence-model-val}, model performance remains robust when trained on these subsplits and evaluated against calculations with tighter convergence thresholds and zero drift.

\subsection{OMat24 model evaluations}

We evaluate the performance of MLIPs trained on the OMat24 dataset and compare against models trained on alternative datasets using the popular Matbench-Discovery benchmark \cite{riebesellFrameworkEvaluateMachine2025}. Both eSEN \cite{fu2025learning} and EquiformerV2 \cite{liao2024equiformerv2} trained on the OMat24 dataset surpass previous state-of-the-art models. Most notably, all five of the currently top-performing models, trained by ourselves or others, on the leaderboard are exclusively trained on the OMat24 dataset. This underscores the critical role of large diverse datasets in advancing MLIP performance.

In addition, we investigate the pervasive issue of systematic underprediction of energy, forces, and phonons---referred to as systematic softening \cite{dengSystematicSofteningUniversal2025}. Our analysis shows that this detrimental effect is dramatically reduced or entirely eliminated across diverse model architectures when trained on the OMat24 dataset. This breakthrough highlights the dataset’s unique ability to enhance model reliability and generalizability to non-equilibrium regimes.

\subsubsection{Model performance on Matbench-Discovery}

The Matbench discovery currently evaluates models based on two prediction tasks: (1) a materials discovery task and (2) a thermal conductivity prediction task. The discovery task predicts thermodynamic stability by comparing a material’s decomposition energy to competing phases \cite{sun2016thermodynamic, bartel_critical_2020, bartel_review_2022}. This requires relaxing atomic positions and the crystal cell, which is more computationally expensive with DFT than with MLIPs. Machine Learning Interatomic Potentials (MLIPs) offer much faster predictions. The recently included thermal conductivity ($\kappa$) prediction task \cite{pota2024thermal, riebesellFrameworkEvaluateMachine2025} requires MLIPs to accurately capture second and third derivatives of the potential energy surface (PES) in order to make accurate predictions of thermal transport.

Table \ref{tab:mbd-noncompliant} summarizes results for models pre-trained on OMat24 and OC20 (metric definitions in Supplementary Table \ref{tab:mbd-metrics-exp}). Because OMat24 uses different DFT settings than the Matbench Discovery leaderboard, we fine-tuned on MPtrj (OMP) or jointly on sAlex and MPtrj (OAM). Pre-training on OC20—despite not being designed for materials modeling—performs strongly after MPtrj fine-tuning, highlighting the value of large-scale non-equilibrium data diversity. Pre-training on OMat24 yields substantially better results across nearly all metrics, clearly outperforming prior methods. The resulting eSEN and equiformerV2 models set the new best performance on Matbench-Discovery  at the time of writing: eSEN achieves an F1 score of 0.925 and an energy-above-hull MAE of 18 meV/atom. The gains from OMat24 are also evident relative to models trained only on MPTrj, with and without denoising (Supplementary Figure \ref{fig:dens-v-data}).

\begin{table*}[!htp]\centering
\caption{Matbench-Discovery benchmark results of non-compliant models on the unique protoype split. Mean absolute error (MAE) and Root mean squared error (RMSE) are in units of eV/atom.\label{tab:mbd-noncompliant}}
\scalebox{0.85}{
\begin{tabular}{l|cccccccc}\toprule
Model &eSEN-30M & eqV2-L & eqV2-M & eqV2-M & eqV2-S & eqV2-S & eqV2-L & eqV2-S \\
Pretrain &OMat &OMat &OMat &OMat & OMat & OMat & OC20 & OC20 \\
Finetune &MPtrj-sAlex &MPtrj-sAlex &MPtrj-sAlex &MPtrj & MPtrj-sAlex & MPtrj & MPtrj & MPtrj \\\midrule
F1 $\uparrow$ &\textbf{0.925} &0.915 &0.917 &0.909 &0.901 &0.89 &0.86 &0.837  \\
DAF $\uparrow$ &6.069 &\textbf{6.113} &6.047 &5.948 &5.902 &5.752 &5.639 &5.392 \\
Precision $\uparrow$ &0.928 &\textbf{0.934} &0.923 &0.909 &0.902 &0.879 &0.862 &0.824 \\
Recall $\uparrow$ &\textbf{0.923} &0.897 &0.910 &0.909 &0.9 &0.901 &0.858 &0.849 \\
Accuracy $\uparrow$ &0.977 &\textbf{0.985} &0.975 &0.973 &0.97 &0.966 &0.957 &0.951 \\ \midrule
TPR $\uparrow$ &\textbf{0.923} &0.897 &0.91 &0.909 &0.9 &0.901 &0.858 &0.849 \\
FPR $\downarrow$ &0.013 &\textbf{0.012} &0.014 &0.017 &0.018 &0.023 &0.025 &0.033 \\
TNR $\uparrow$ &0.987 &\textbf{0.988} &0.986 &0.983 &0.982 &0.977 &0.975 &0.967 \\
FNR $\downarrow$ &\textbf{0.077} &0.103 &0.090 &0.091 &0.1 &0.099 &0.142 &0.151 \\ \midrule
MAE $\downarrow$ &\textbf{18} &19 &20 &21 &24 &26 &29 &33 \\
RMSE $\downarrow$ &\textbf{67} &71 &72 &72 &80 &81 &78 &80 \\
R2 $\uparrow$ &\textbf{0.866} &0.852 &0.848 &0.849 &0.811 &0.807 &0.823 &0.810 \\
\bottomrule
\end{tabular}}
\end{table*}

\begin{figure*}[ht!]
    \centering
    \includegraphics[width=0.7\textwidth]{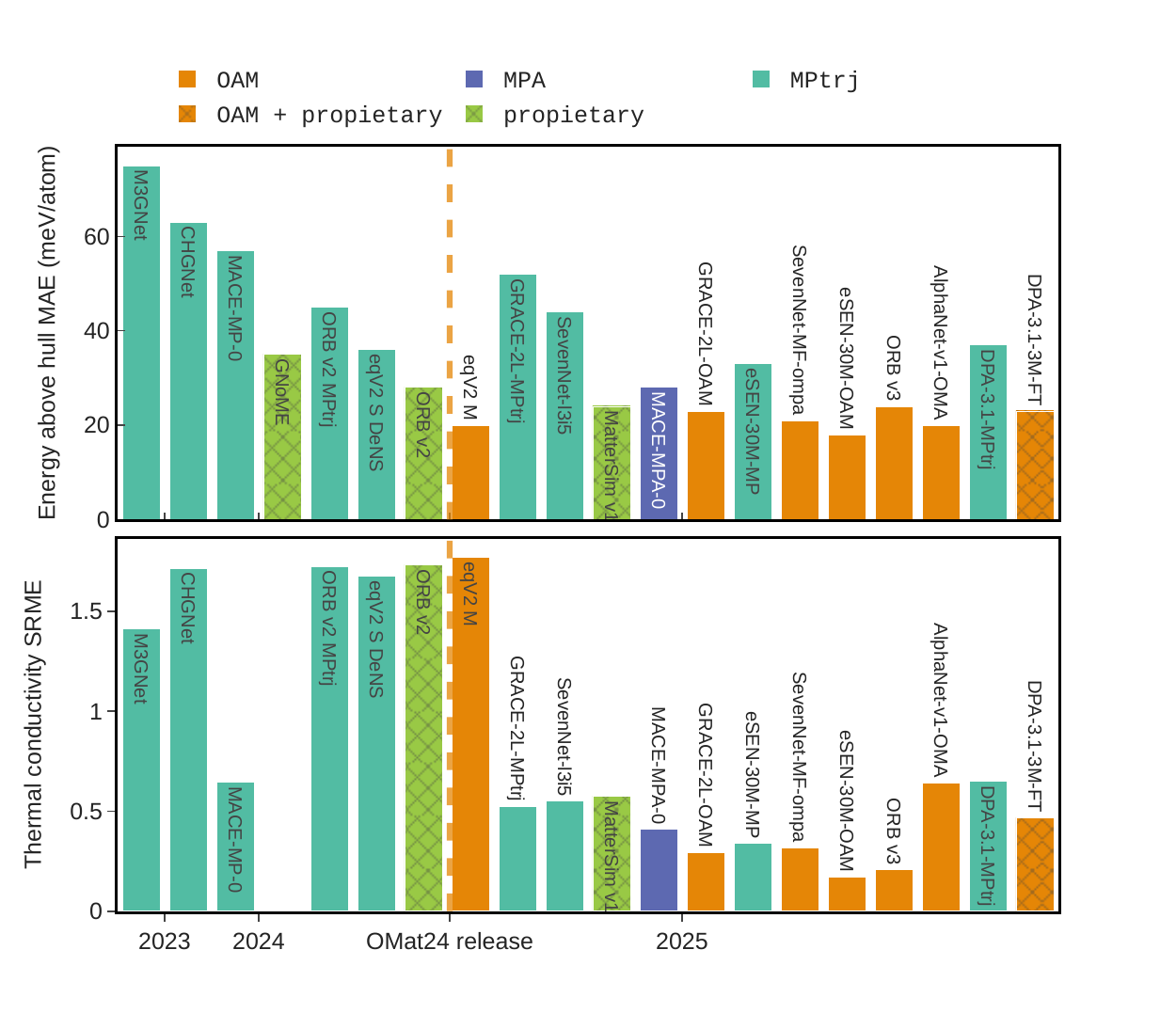}
    \caption{\textbf{Chronological performance improvements of MLIP models.} Energy above hull mean absolute error and thermal conductivity symmetric relative mean error ($\kappa_{SRME}$) of models on Matbench-Discovery \cite{riebesellFrameworkEvaluateMachine2025} in chronological release order. The majority top performing models are trained on the OMat24 dataset and finetuned on MPtrj \cite{deng_chgnet_2023} and sAlex \cite{schmidt_machine_2023}, labeled as OAM.} 
    \label{fig:oam-model-accuracy}
\end{figure*}

More importantly, the substantial improvements from models trained using OMat24 extend across architectures beyond eSEN and equiformerV2. Figure \ref{fig:oam-model-accuracy} displays the mean absolute error for energy above hull and the symmetric relative mean error (SRME) for thermal conductivity, presented in chronological order for models on the Matbench-Discovery leaderboard\cite{riebesellFrameworkEvaluateMachine2025} from its inception to early 2025, six months after the release of the OMat24 preprint \url{https://arxiv.org/abs/2410.12771}.

The five top performing models trained using OMat24 achieve unprecedented performance in the discovery tasks, with many achieving energy above hull prediction errors near 20 meV/atom. Additionally, we observe that all \emph{energy conserving} models trained using OMat24 exhibit the lowest thermal conductivity prediction errors and consistently top the overall Matbench-Discovery leaderboard. We note that direct force models, such as equiformerV2 and ORB v2, have been found to perform poorly in phonon prediction tasks \cite{loewUniversalMachineLearning2025, winesCHIPSFFEvaluatingUniversal2025}, as evidenced by their high thermal conductivity SRME.

\subsubsection{Model evaluation of systematic softening}

A systematic underprediction of energy, forces, and phonons has been observed across MLIP architectures trained on smaller datasets of relaxation trajectories, such as MPtrj \cite{dengSystematicSofteningUniversal2025}. This phenomenon, known as systematic softening, is attributed to dataset distribution bias and a lack of non-equilibrium structures.

Building on the work of Deng et al. \cite{dengSystematicSofteningUniversal2025}, we quantify systematic softening at three levels based on the order of potential energy derivatives used for prediction.

\begin{enumerate}
    \item \textbf{Zeroth order softening} quantifies the systematic underprediction of energy in non-equilibrium structures, such as those with defects, surfaces, and interfaces. We measure this as the difference between the target and predicted energy per atom. A negative bias indicates zeroth-order softening. 
    \item \textbf{First order softening} quantifies the systematic underprediction of forces in non-equilibrium structures. We assess this using the previously proposed \emph{softening scale} \cite{dengSystematicSofteningUniversal2025}, which is the slope of the least squares fit to the parity plot of target forces versus predicted forces. A bias towards values smaller than one indicates first-order softening. 
    \item \textbf{Second order softening} quantifies the systematic underprediction of phonon vibrational frequencies. We measure this as the ratio between the maximum predicted and maximum target phonon frequency. A bias towards values smaller than one indicates second-order softening.
\end{enumerate}

We evaluate zeroth-order and first-order softening using the WBM high energy states dataset \cite{dengSystematicSofteningUniversal2025}, which comprises approximately ten thousand high-temperature MD samples from randomly selected WBM structures \cite{wang_predicting_2021}. We estimate second-order softening using the PBE MDR phonon dataset \cite{loewUniversalMachineLearning2025}. All calculations in these two datsets use the PBE functional and the same pseudopotentials; the MDR phonon calculations differ primarily in their more stringent force relaxation and electronic energy convergence criteria \cite{loewUniversalMachineLearning2025}. We note that the PBE MDR phonon dataset excludes materials requiring +U corrections per the Materials Project protocol \cite{horton2025accelereated}.

\begin{figure*}[ht!]
    \centering
    \includegraphics[width=\textwidth]{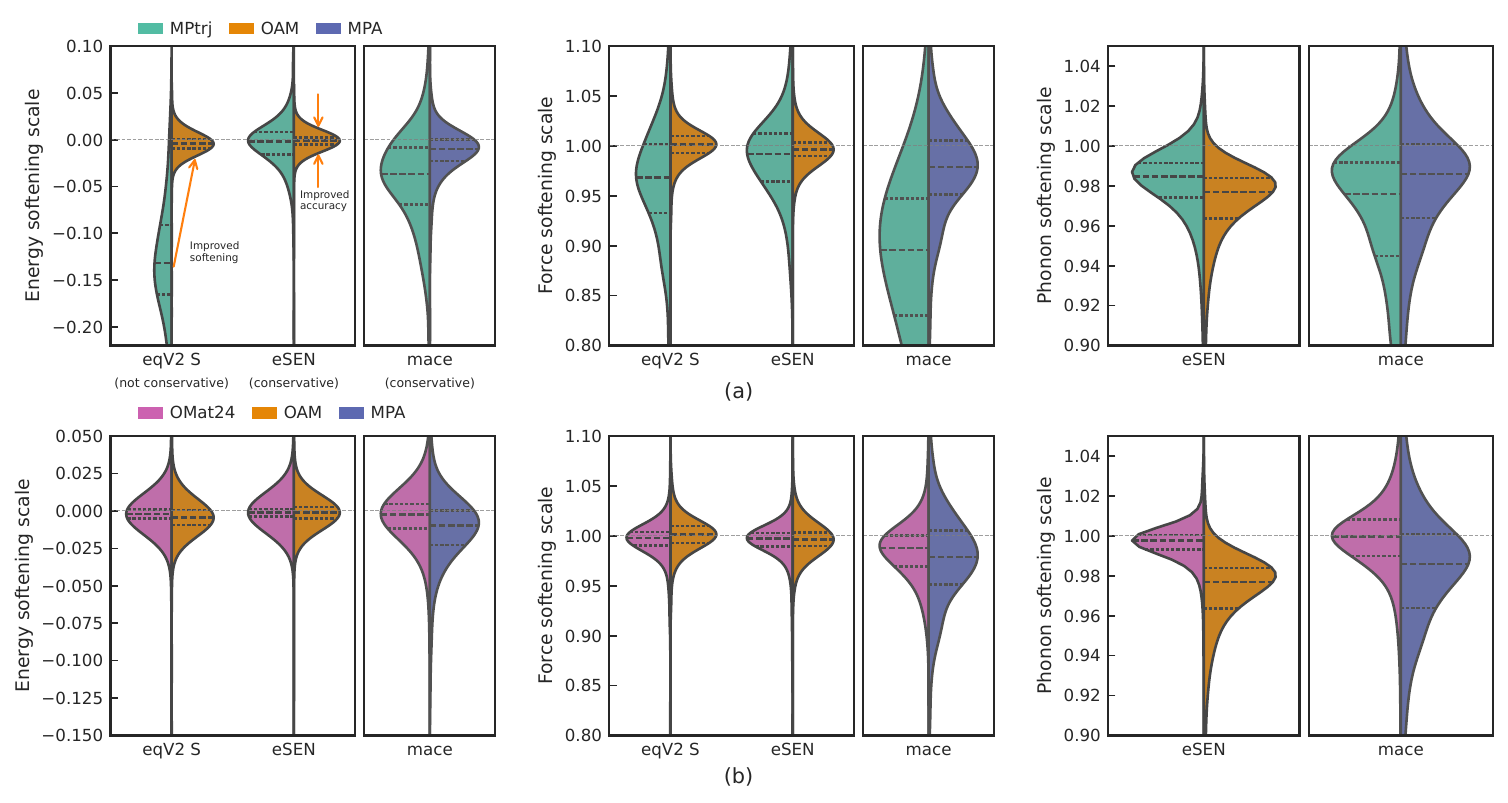}
    \caption{\textbf{Reduction of systematic softening across models using OMat24 training data} Zeroth order (energy), first order (forces) and second order (phonon) systematic softening distribution shifts between (a) MPTrj, OAM (OMat24 + MP/sAlex finetuning) and MPA (MP/sAlex) trained models and (b) OMat24, OAM, and MPA models. Violin plots show the distribution of the energy and force softening scale of model predictions on high energy WBM set \cite{dengSystematicSofteningUniversal2025}, and phonon softening of conservative model predictions on the recalculated MDR phonon dataset \cite{loewUniversalMachineLearning2025}. Models are labeled \emph{conservative} if forces are predicted as energy gradients, and \emph{not conservative} otherwise.}
    \label{fig:softening-omat}
\end{figure*}

Figure \ref{fig:softening-omat} shows violin plots quantifying (a) zeroth-order, (b) first-order, and (c) second-order systematic softening for models trained on MPtrj only, pretrained on OMat24 and finetuned on MPtrj and sAlex (OAM), and trained on OMat24 only. MACE models trained on MPtrj and MPtrj + sAlex without OMat24 pretraining are included as baselines; results for additional third-party models appear in Supplementary Figure \ref{fig:softening-all}. OAM models show a notable reduction in zeroth- and first-order softening compared to MPtrj-only models, with the eSEN and equiformerV2 architectures nearly eliminating it entirely. The MACE model trained on MPtrj and sAlex only (MPA) also exhibits reduced softening relative to the MPtrj-only model, though to a lesser extent than the OAM models.

The second-order (phonon) softening results in Figure \ref{fig:softening-omat}a show a regression for the eSEN OAM model, which exhibits more softening than its MPtrj counterpart. In contrast, the GRACE and SevenNet OAM models show moderate reductions in softening (Supplementary Figure \ref{fig:softening-all}). To understand this regression, we examine models trained on OMat24 alone without fine-tuning (Figure \ref{fig:softening-omat}b). The eSEN OMat24-only model does not exhibit second-order softening, highlighting the benefit of the dataset's diversity and breadth. Indeed, models trained solely on OMat24 show less softening of all orders across all architectures, suggesting that fine-tuning with relaxation data reintroduces some systematic softening.

\section{Discussion}

The OMat24 dataset is a major step forward for materials science, providing a large, diverse set of PBE-DFT structures for training ML interatomic potentials. Its breadth enables models to capture complex material behavior and considerably improve property prediction accuracy. While OMat24 samples a broad spectrum of non-equilibrium configurations, it does not explicitly include transition-state frames or highly off-equilibrium regions such as bond-breaking or large coordination changes. Most existing benchmarks focus on equilibrium or near-equilibrium properties; nevertheless, models trained on OMat24 have shown strong extrapolation capabilities, successfully predicting properties for states not directly represented in the dataset—including defects, grain boundaries, adsorption, and cleavage energies \cite{shuangUniversalMachineLearning2025, moonCatBenchFrameworkBenchmarking2025, mehdizadehSurfaceStabilityModeling2025}. Ongoing development of such benchmarks remains essential for identifying model limitations and guiding future dataset generation efforts.

Furthermore, discrepancies between the PBE functional and experimental results can limit the reliability of predictions for real-world applications. Users should be mindful of the inherent limitations of the PBE(+U) functional and the dataset’s emphasis on bulk, defect-free structures. In particular, PBE and related GGA functionals often struggle to accurately capture strong on-site Coulomb interactions in d and f orbitals, which may lead to deviations from experimental observations. \cite{rezaeiEvaluatingSCANR$^2$SCAN2025}.

Developing training sets and benchmarks using more accurate DFT functionals such as SCAN and r2SCAN could help in starting to overcome these challenges \cite{kingsbury2022performance}. As of release v2022.10.28 \cite{MP-release}, the Materials Project began including r2SCAN calculations for a portion of its dataset. Additionaly, MatPES, the first r2SCAN MLIP training dataset containing 400K structures \cite{kaplanFoundationalPotentialEnergy2025a} was released shortly after the release of OMat24. Large-scale datasets similar to OMat24 but with SCAN, r2SCAN or higher fidelity functionals along with additional physically motivated sampling strategies, such as using active learning cycles, molecular dynamics, and metadynamics to efficiently sample and optimize configurations beyond local minima are obvious next steps.

Efficiently building models with higher-fidelity methods is also possible through multi-fidelity learning, delta learning, or fine-tuning. These approaches can leverage the extensive low-level force data available in OMat24, supplemented by a smaller set of high-level energies, allowing MLIPs to achieve hybrid or even CCSD(T)-level accuracy with reduced computational cost \cite{messerlyMultifidelityLearningInteratomic2025a, ikedaMachinelearningInteratomicPotentials2025, jacobs2025practical}. By using OMat24’s elemental diversity and non-equilibrium sampling as a pretraining foundation, researchers can accelerate the development of high-accuracy potentials for applications where functional errors or surface and defect physics are critical.

We anticipate that the release of OMat24 will facilitate more data-efficient modeling efforts. For instance, the MatPES project demonstrated the value of maximizing information content in smaller datasets \cite{kaplan2020scaling}, and recent work has focused on effective sampling and dataset compression strategies \cite{liExploitingRedundancyLarge2023, yuMaximizingEfficiencyDataset2025}. OMat24 provides a robust platform to further investigate sampling and compression techniques aimed at identifying non-redundant physical information, ultimately enabling the development of state-of-the-art models trained on minimal data.

The OMat24 dataset has quickly become a pivotal resource in developing highly accurate MLIPs, greatly reducing systematic softening biases. Its comprehensive coverage enable the creation of models that not only excel in predictive tasks but also provide a reliable foundation for future advancements in materials science. Similar to how this work builds upon open research efforts, like the Materials Project \cite{jain2013commentary} and the Alexandria datasets \cite{schmidt_machine_2023}, we hope this work enables the research comunity to drive further advancements in the capabilities of machine learning models for materials science.

\section{Methods}

\subsection{OMat24 structure generation and sampling}

Input structure generation for the OMat24 dataset consists of three different processes intended to obtain diverse non-equilibrium structures: Boltzmann sampling of rattled (random Gaussian perturbations of atomic positions) structures, ab initio molecular dynamics (AIMD), and relaxations of rattled structures. These methods were used to increase the diversity of sampled configurations, similar to prior large dataset efforts \cite{merchant_scaling_2023, yang2024mattersim}. 

In all three approaches, initial structures were obtained by randomly sampling the relaxed structures in the Alexandria PBE bulk materials dataset (3D compounds) \cite{schmidt_machine_2023}. The Alexandria dataset was chosen as a starting point because it was the largest openly available DFT dataset of equilibrium and near equilibrium structures ($\sim$4.5 million materials). By randomly sampling relaxed structures from the Alexandria dataset we are able to cover a wide elemental composition diversity. Additionally, using Alexandria relaxed structures as a starting point prevents generating structures too far from equilibrium that could result in DFT convergence errors, or unphysical starting configurations. The details for the three processes are as follows:
\begin{itemize}
    \item[] \textbf{Rattled Boltzmann sampling}: For each randomly sampled Alexandria structure we generated 500 candidate non-equilibrium structures by scaling the unit cell to contain at least 10 atoms. Atomic positions were then rattled with displacements sampled from a Gaussian distribution of $\mu=0$\r{A} and $\sigma=0.5$\r{A}. Unit cells were also deformed isotropically and anisotropically from a Gaussian distribution of $\mu=1\%$ and $\sigma=5\%$. For each set of 500 candidate structures we selected 5 of them from a Boltzmann-like distribution based on total energies predicted by an EquiformerV2 model trained on the MPtrj dataset. The sampling procedure was done using three different sampling temperatures: 300K, 500K and 1000K.
    \item[] \textbf{Ab-Initio Molecular Dynamics (AIMD)}: Short-length (50 ionic steps) ab-initio molecular dynamics were carried out starting from randomly sampled relaxed structures in the Alexandria dataset. Structures were recorded from constant temperature and volume (NVT), and constant temperature and pressure (NPT) AIMD trajectories at temperatures of 1000K and 3000K. Unit cells were scaled to contain at least 50 atoms for structures sampled at 3000K.
    \item[] \textbf{Rattled relaxation}: Relaxed structures from Alexandria were selected at random, rattled (both atomic positions and unit cell), and re-relaxed. Atomic displacements were sampled from a Gaussian distribution of $\mu=0$\r{A} and $\sigma=0.2$\r{A}. Similarly, isotropic and anisotropic cell deformations were sampled with a $\mu=1\%$ and $\sigma=4\%$. All structures along the relaxation trajectory were included in the dataset.
\end{itemize}

We note that these strategies were chosen to increase diversity, but there are many possible sampling strategies that could be used to maximize information content \cite{schwalbe2021differentiable, li2023exploiting}, and we expect these strategies to be useful for future sampling efforts. Active learning\cite{yang2024mattersim, merchant_scaling_2023} sampling strategies have the potential to further enhance these approaches but it remains unclear how they compare to random baselines when considering large scale dataset sizes.

\subsection{OMat24 DFT calculation settings and details} \label{sec:omat-dft-settings}
DFT calculations generally followed Material Project default settings \cite{jain2013commentary} with some important exceptions. The calculations in this work have been performed using the ab-initio total-energy and molecular-dynamics package VASP (Vienna ab-initio simulation package) developed at the Institut f{\"u}r Materialphysik of the Universi{\"a}t Wien \cite{Kresse1994, Kresse1996a} with periodic boundary conditions and the projector augmented wave (PAW) pseudopotentials \cite{kresse_ultrasoft_1999}. Exchange and correlation effects were calculated using the generalized gradient approximation and the Perdew-Burke-Ernzerhof (PBE) with Hubbard U corrections for oxide and fluoride materials containing Co, Cr, Fe, Mn, Mo, Ni, V, or W, following Materials Project defaults \cite{jain2013commentary}. 

VASP input sets were generated using the \textsc{MPRelaxSet} class defined in the \textsc{pymatgen}\cite{ong_python_2013} library with the following modifications to account for recent updates and changes in the underlying algorithms and pseudopotentials:
\begin{enumerate}
    \item Version 54 of pseudopoentials provided by VASP were used, rather than the legacy PBE MPRelaxSet defaults \cite{jain2013commentary}. The Yb\_3 and W\_sv pseudopotentials were used for Yb and W to account for changes between version 52 and 54 of VASP PBE pseudopotentials \cite{pymatgen-issue-2968, pymatgen-issue-3016}. 
    \item  All calculations were done with the \textsc{ALGO} flag set to ``Normal''. 
\end{enumerate}
A list of all differing pseudopotentials POTCAR symbols and generation dates is given in the Supplementatary Section \ref{sec:psuedo-diffs}.

Relaxations were conducted with the \textsc{MPRelax} set defaults of VASP input flags.  All AIMD calculations were carried out for 50 steps at a time-step of 2 femtoseconds. We note that 2 fs is a large time-step for typical AIMD simulations, especially for hydrogen-containing materials, but was chosen as the goal was to sample diverse configurations rather than perfectly integrate the trajectories. Finally, for static calculations resulting from inputs generated using rattling and Boltzmann sampling, only the NSW and IBRION flags where updated as appropriate for single point calculations.

\subsection{OMat24 MLIP training strategies}

We used the OMat24 dataset along with the MPtrj \cite{deng_chgnet_2023} and Alexandria \cite{schmidt_machine_2023} datasets to train GNN MLIPs. Since similar structures exist in the Alexandria and the WBM dataset used for testing, we subsampled the Alexandria dataset for training to ensure there was no leakage between the training and testing datasets. The new subset of Alexandria (sAlexandria) was created by removing all trajectories in which any structure matched a structure in the WBM initial and relaxed structures using structure prototype labels \cite{mehl_aflow_2017}. Next, we reduced the size of the Alexandria dataset by removing all structures with energies > 0 eV, forces norm > 50 eV/\r{A}, and stress > 80 GPa. Finally, we only sampled structures in the remaining trajectories that had a energy difference greater than 10 meV/atom. The resulting datasets used for training and validation had 10 million and 500 thousand structures respectively.

We train and evaluate EquiformerV2 models \cite{liao2024equiformerv2} as it is currently one of the top performing direct force models on the OC20 \cite{oc20}, OC22 \cite{tran_open_2023} and ODAC23 \cite{sriram_open_2024} benchmarks. We also train and evaluate eSEN models, since it is an energy conserving architecture which has shown SOTA performance for phonon and thermal conductivity calculations \cite{fu2025learning}. The models are trained to predict energy, forces and stress given an input structure. We explored three strategies:
\begin{enumerate}
    \item EquiformerV2/eSEN models trained solely on the OMat24 dataset, with and without denoising augmentation objectives. 
    \item EquiformerV2/eSEN models trained solely on the MPtrj dataset, with and without denoising augmentation objectives, useful for direct comparison on the Matbench Discovery leaderboard (denoted ``compliant`` models).
    \item EquiformerV2/eSEN models from (1) or OC20 checkpoints further fine-tuned on either the MPtrj or sAlexandria datasets, leading to the highest performing models for the Matbench Discovery leaderboard (denoted ``non-compliant``). 
\end{enumerate}

Additional model specifications, training information, and parameters are given in Supplementary Section \ref{sec:supp-training}.

\subsection{MLIP evaluations of energy, force and phonon softening}

To evaluate systematic softening in model predictions, we predicted energy and forces for the \~ 1000 structures in the WBM high energy states dataset \cite{dengSystematicSofteningUniversal2025}, and phonon frequencies for approximately 10,000 structures in the PBE MDR phonon dataset \cite{loewUniversalMachineLearning2025}. The zeroth order (energy) softening was computed as the difference in the predicted energy per atom with the DFT-calculated energy per atom. First order (force) softening is computed as the slope of the least squares fit of the predicted forces and the target forces \cite{dengSystematicSofteningUniversal2025}. Finally, second order (phonon) softening was computed as the ratio of the maximum predicted phonon frequency and the maximum DFT target phonon frequency.

We computed energy and force softening of all OAM models using the DFT energy and force values in the WBM high energy dataset \cite{dengSystematicSofteningUniversal2025} which are computed with DFT settings that match Materials Project and Matbench-discovery settings. Additionally, we ran single-point DFT calculations using the OMat24 settings described in Section \ref{sec:omat-dft-settings} in order to compute energy and force labels to evaluate softening of OMat24 trained models.

Energy and force softening calculations are carried out each architecture (eSEN, equiformerV2, GRACE, SevenNet and MACE) by running model predictions using each of their \textsc{ASE} compatible calculator classes. Phonon softening is similarly calculated by running model inference via \textsc{ASE} calculators following the following protocol \cite{loewUniversalMachineLearning2025}:
\begin{enumerate}
    \item Run a structure relaxation of the primitive cell with a convergence force threshold of 5 meV/\AA using the FIRE optimizer with \textsc{FrechetCellFilter} implemented in \textsc{ASE} \cite{ase-paper}.
    \item Calculate force constants using the finite displacement supercell method implemented in \textsc{Phonopy} \cite{phonopy-phono3py-JPSJ} using a displacement of 0.01 \AA. The supercell for all structures were taken directly from the PBE MDR dataset.
    \item Calculate the maximumum phonon frequency from phonon frequencies calculated at q-points commensurate with the supercell.
\end{enumerate}
\clearpage





\section*{Data Availability}

The OMat24 and subsampled Alexandria (sAlex) datasets, as well as the reference compounds and fitted MP2020-style Compatiblity corrections are available for download via Hugging Face \cite{omat24}. VASP input generation files necessary to reproduce DFT calculations can be found in the \textsc{fairchem-data-omat} repository \cite{fairchem-omat-data}. Source data for Figures \ref{fig:omat-overview}-\ref{fig:softening-omat} is available with this manuscript.

\section*{Code Availability}

All model training and architecture code is available on the \textsc{FAIRChem} Github \cite{fairchemrepo}. equiformerV2 and eSEN pretrained model checkpoints can be downloaded from Hugging Face \cite{omat24_models}. Pretrained models are compatible with \textsc{fairchem-core} 1.10 \cite{fairchemrepo}. 
\section*{Acknowledgements}

We acknowledge helpful discussions with Miguel Marques (Ruhr University Bochum), Janosh Riebesell, Shyue Ping Ong (UCSD), Yi-Lun Liao (MIT), Tess Smidt (MIT), and Bowen Deng (U.C. Berkeley).  The authors acknowledge valuable assistance from Kyle Michel and Lowik Chanussot (FAIR at Meta) in enabling and scaling the simulation datasets on Meta compute resources. We generated several of our figures using \textsc{crystal-toolkit}\cite{horton2023crystal} and \textsc{pymatviz}\cite{riebesell_pymatviz_2022}.

\section*{Author Contributions}
L.B.L., M.S., M.U., C.L.Z, and Z.U conceived, designed and planned the project. L.B.L. and M.S. generated input structures, ran DFT and processed all DFT calculations. L.B.L., M.S., X.F. and B.M.W. optimized hyperparameters and trained models. M.D. and M.G. implemented training and model inference infrastructure. A.R., M.U., and Z.U. supervised the work. L.B.L., M.S., X.F., B.M.W., C.L.Z, and Z.U wrote the paper. All authors edited the paper.

\section*{Competing Interests}
Authors declare no competing interests.
\bibliographystyle{unsrt}
\bibliography{refs}

@softmisc{fairchem-omat-data,
  doi = {10.5281/ZENODO.19687634},
  url = {https://zenodo.org/doi/10.5281/zenodo.19687634},
  author = {Barroso-Luque, Luis and Shuaibi, Muhammed and Ulissi, Zachary},
  title = {facebookresearch/fairchem: fairchem\_data\_omat-0.2.0},
  publisher = {Zenodo},
  year = {2025},
  copyright = {Creative Commons Attribution 4.0 International}
}

@softmisc{fairchemrepo,
  doi = {10.5281/ZENODO.19687080},
  url = {https://zenodo.org/doi/10.5281/zenodo.19687080},
  author = {{Meta Fundamental AI Research and Collaborators}},
  title = {facebookresearch/fairchem: fairchem\_core-1.10.0},
  publisher = {Zenodo},
  year = {2025},
  copyright = {Creative Commons Attribution 4.0 International}
}

@misc{omat24,
	author       = { {Meta Fundamental AI Research} },
	title        = { OMAT24 },
	year         = 2026,
	url          = { https://huggingface.co/facebook/OMAT24 },
	doi          = { 10.57967/hf/8227 },
	publisher    = { Hugging Face }
}

@misc{omat24_models,
	author       = { {Meta Fundamental AI Research} },
	title        = { OMAT24 },
	year         = 2026,
	url          = { https://huggingface.co/datasets/facebook/OMAT24 },
	doi          = { 10.57967/hf/8434 },
	publisher    = { Hugging Face }
}

@article{ikedaMachinelearningInteratomicPotentials2025,
  title={Machine-learning interatomic potentials achieving CCSD (T) accuracy for systems with extended covalent networks and van der Waals interactions},
  author={Ikeda, Yuji and Forslund, Axel and Kumar, Pranav and Ou, Yongliang and Jung, Jong Hyun and Koohn, Andreas and Grabowski, Blazej},
journal = {Journal of Chemical Theory and Computation},
volume = {22},
number = {6},
pages = {2739-2756},
year = {2026},
doi = {10.1021/acs.jctc.5c02045},
note ={PMID: 41774831},
URL = { https://doi.org/10.1021/acs.jctc.5c02045}
}

@article{kunerMPALOER2SCANDataset2025,
  title = {{{MP-ALOE}}: An {{r2SCAN}} Dataset for Universal Machine Learning Interatomic Potentials},
  author = {Kuner, Matthew C. and Kaplan, Aaron D. and Persson, Kristin A. and Asta, Mark and Chrzan, Daryl C.},
  year = 2025,
  month = nov,
  journal = {npj Computational Materials},
  volume = {11},
  number = {1},
  pages = {352},
  publisher = {Nature Publishing Group},
  issn = {2057-3960},
  doi = {10.1038/s41524-025-01834-9},
  copyright = {2025 The Author(s)}
}

@article{mehdizadehSurfaceStabilityModeling2025,
  title = {Surface Stability Modeling with Universal Machine Learning Interatomic Potentials: A Comprehensive Cleavage Energy Benchmarking Study},
  author = {Mehdizadeh, Ardavan and Schindler, Peter},
  year = 2025,
  month = nov,
  journal = {AI for Science},
  volume = {1},
  number = {2},
  pages = {025002},
  publisher = {IOP Publishing},
  issn = {3050-287X},
  doi = {10.1088/3050-287X/ae1408}
}

@article{messerlyMultifidelityLearningInteratomic2025a,
  title = {Multi-Fidelity Learning for Interatomic Potentials: Low-Level Forces and High-Level Energies Are All You Need*},
  author = {Messerly, Mitchell and Matin, Sakib and Allen, Alice E A and Nebgen, Benjamin and Barros, Kipton and Smith, Justin S and Lubbers, Nicholas and Messerly, Richard},
  year = 2025,
  month = sep,
  journal = {Machine Learning: Science and Technology},
  volume = {6},
  number = {3},
  pages = {035066},
  publisher = {IOP Publishing},
  issn = {2632-2153},
  doi = {10.1088/2632-2153/ae040b}
}

@article{moonCatBenchFrameworkBenchmarking2025,
  title = {{{CatBench}} Framework for Benchmarking Machine Learning Interatomic Potentials in Adsorption Energy Predictions for Heterogeneous Catalysis},
  author = {Moon, Jinuk and Jeon, Uchan and Choung, Seokhyun and Han, Jeong Woo},
  year = 2025,
  month = dec,
  journal = {Cell Reports Physical Science},
  volume = {6},
  number = {12},
  publisher = {Elsevier},
  issn = {2666-3864},
  doi = {10.1016/j.xcrp.2025.102968}
}

@article{rezaeiEvaluatingSCANR$^2$SCAN2025,
  title={Evaluating SCAN and r 2 SCAN meta-GGA functionals for predicting transition temperatures in antiferromagnetic materials},
  author={Rezaei, Nafise and Alaei, Mojtaba and Oganov, Artem R},
  journal={Physical Review B},
  volume={111},
  number={14},
  pages={144406},
  year={2025},
  publisher={APS}
}

@article{shuangUniversalMachineLearning2025,
  title = {Universal Machine Learning Interatomic Potentials Poised to Supplant {{DFT}} in Modeling General Defects in Metals and Random Alloys},
  author = {Shuang, Fei and Wei, Zixiong and Liu, Kai and Gao, Wei and Dey, Poulumi},
  year = 2025,
  month = jul,
  journal = {Machine Learning: Science and Technology},
  volume = {6},
  number = {3},
  pages = {030501},
  publisher = {IOP Publishing},
  issn = {2632-2153},
  doi = {10.1088/2632-2153/adea2d}
}

@article{jacobs2025practical,
  title={A practical guide to machine learning interatomic potentials--Status and future},
  author={Jacobs, Ryan and Morgan, Dane and Attarian, Siamak and Meng, Jun and Shen, Chen and Wu, Zhenghao and Xie, Clare Yijia and Yang, Julia H and Artrith, Nongnuch and Blaiszik, Ben and others},
  journal={Current Opinion in Solid State and Materials Science},
  volume={35},
  pages={101214},
  year={2025},
  publisher={Elsevier}
}

@article{sanspeurCircumventingDataImbalance2024,
  title = {Circumventing Data Imbalance in Magnetic Ground State Data for Magnetic Moment Predictions},
  author = {Sanspeur, Rohan Yuri and Kitchin, John R},
  year = 2024,
  month = feb,
  journal = {Machine Learning: Science and Technology},
  volume = {5},
  number = {1},
  pages = {015023},
  publisher = {IOP Publishing},
  issn = {2632-2153},
  doi = {10.1088/2632-2153/ad23fb}
}

@article{xuSpininformedUniversalGraph2025,
  title = {Spin-Informed Universal Graph Neural Networks for Simulating Magnetic Ordering},
  author = {Xu, Wenbin and Sanspeur, Rohan Yuri and Kolluru, Adeesh and Deng, Bowen and Harrington, Peter and Farrell, Steven and Reuter, Karsten and Kitchin, John R.},
  year = 2025,
  month = jul,
  journal = {Proceedings of the National Academy of Sciences},
  volume = {122},
  number = {27},
  pages = {e2422973122},
  publisher = {Proceedings of the National Academy of Sciences},
  doi = {10.1073/pnas.2422973122}
}

@article{liExploitingRedundancyLarge2023,
  title = {Exploiting Redundancy in Large Materials Datasets for Efficient Machine Learning with Less Data},
  author = {Li, Kangming and Persaud, Daniel and Choudhary, Kamal and DeCost, Brian and Greenwood, Michael and {Hattrick-Simpers}, Jason},
  year = 2023,
  month = nov,
  journal = {Nature Communications},
  volume = {14},
  number = {1},
  pages = {7283},
  publisher = {Nature Publishing Group},
  issn = {2041-1723},
  doi = {10.1038/s41467-023-42992-y},
  copyright = {2023 His Majesty the King in Right of Canada as represented by the Minister of Natural Resources}
}

@misc{yuMaximizingEfficiencyDataset2025,
  title = {Maximizing {{Efficiency}} of {{Dataset Compression}} for {{Machine Learning Potentials With Information Theory}}},
  author = {Yu, Benjamin and Lordi, Vincenzo and {Schwalbe-Koda}, Daniel},
  year = 2025,
  month = nov,
  number = {arXiv:2511.10561},
  eprint = {2511.10561},
  primaryclass = {cs},
  publisher = {arXiv},
  doi = {10.48550/arXiv.2511.10561},
  archiveprefix = {arXiv}
}

@misc{sahoo2025opencatalyst2025oc25,
      title={The Open Catalyst 2025 (OC25) Dataset and Models for Solid-Liquid Interfaces}, 
      author={Sushree Jagriti Sahoo and Mikael Maraschin and Daniel S. Levine and Zachary Ulissi and C. Lawrence Zitnick and Joel B Varley and Joseph A. Gauthier and Nitish Govindarajan and Muhammed Shuaibi},
      year={2025},
      eprint={2509.17862},
      archivePrefix={arXiv},
      primaryClass={cond-mat.mtrl-sci},
      url={https://arxiv.org/abs/2509.17862}, 
}

@article{kuryla2025accuratedftforces,
    author = {Kuryla, Domantas and Berger, Fabian and Csányi, Gábor and Michaelides, Angelos},
    title = {How accurate are DFT forces? Unexpectedly large uncertainties in molecular datasets},
    journal = {The Journal of Chemical Physics},
    volume = {163},
    number = {22},
    pages = {224313},
    year = {2025},
    month = {12},
    issn = {0021-9606},
    doi = {10.1063/5.0296997},
    url = {https://doi.org/10.1063/5.0296997},
}

@article{horton2025accelereated,
  title = {Accelerated Data-Driven Materials Science with the {{Materials Project}}},
  author = {Horton, Matthew K. and Huck, Patrick and Yang, Ruo Xi and Munro, Jason M. and Dwaraknath, Shyam and Ganose, Alex M. and Kingsbury, Ryan S. and Wen, Mingjian and Shen, Jimmy X. and Mathis, Tyler S. and Kaplan, Aaron D. and Berket, Karlo and Riebesell, Janosh and George, Janine and Rosen, Andrew S. and {Spotte-Smith}, Evan W. C. and McDermott, Matthew J. and Cohen, Orion A. and Dunn, Alex and Kuner, Matthew C. and Rignanese, Gian-Marco and Petretto, Guido and Waroquiers, David and Griffin, Sinead M. and Neaton, Jeffrey B. and Chrzan, Daryl C. and Asta, Mark and Hautier, Geoffroy and Cholia, Shreyas and Ceder, Gerbrand and Ong, Shyue Ping and Jain, Anubhav and Persson, Kristin A.},
  year = {2025},
  month = oct,
  journal = {Nature Materials},
  volume = {24},
  number = {10},
  pages = {1522--1532},
  publisher = {Nature Publishing Group},
  issn = {1476-4660},
  doi = {10.1038/s41563-025-02272-0},
  copyright = {2025 Springer Nature Limited}
}

@misc{wang2025graph,
      title={A Graph Neural Network for the Era of Large Atomistic Models}, 
      author={Duo Zhang and Anyang Peng and Chun Cai and Wentao Li and Yuanchang Zhou and Jinzhe Zeng and Mingyu Guo and Chengqian Zhang and Bowen Li and Hong Jiang and Tong Zhu and Weile Jia and Linfeng Zhang and Han Wang},
      year={2026},
      eprint={2506.01686},
      archivePrefix={arXiv},
      url={https://arxiv.org/abs/2506.01686}, 
}

@inproceedings{krizhevsky2012imagenet,
 author = {Krizhevsky, Alex and Sutskever, Ilya and Hinton, Geoffrey E},
 booktitle = {Advances in Neural Information Processing Systems},
 editor = {F. Pereira and C.J. Burges and L. Bottou and K. Weinberger},
 pages = {},
 publisher = {Curran Associates, Inc.},
 title = {ImageNet Classification with Deep Convolutional Neural Networks},
 url = {https://proceedings.neurips.cc/paper_files/paper/2012/file/c399862d3b9d6b76c8436e924a68c45b-Paper.pdf},
 volume = {25},
 year = {2012}
}

@article{kaplan2020scaling,
  title={Scaling laws for neural language models},
  author={Kaplan, Jared and McCandlish, Sam and Henighan, Tom and Brown, Tom B and Chess, Benjamin and Child, Rewon and Gray, Scott and Radford, Alec and Wu, Jeffrey and Amodei, Dario},
  journal={arXiv preprint arXiv:2001.08361},
  year={2020}
}

@misc{kaplanFoundationalPotentialEnergy2025a,
  title = {A {{Foundational Potential Energy Surface Dataset}} for {{Materials}}},
  author = {Kaplan, Aaron D. and Liu, Runze and Qi, Ji and Ko, Tsz Wai and Deng, Bowen and Riebesell, Janosh and Ceder, Gerbrand and Persson, Kristin A. and Ong, Shyue Ping},
  year = {2025},
  month = mar,
  number = {arXiv:2503.04070},
  eprint = {2503.04070},
  primaryclass = {cond-mat},
  publisher = {arXiv},
  doi = {10.48550/arXiv.2503.04070},
  archiveprefix = {arXiv}
}

@article{winesCHIPSFFEvaluatingUniversal2025,
  title={Chips-ff: Evaluating universal machine learning force fields for material properties},
  author={Wines, Daniel and Choudhary, Kamal},
  journal={ACS Materials Letters},
  volume={7},
  number={6},
  pages={2105--2114},
  year={2025},
  publisher={ACS Publications}
}

@article{pota2024thermal,
  title={Thermal conductivity predictions with foundation atomistic models},
  author={P{\'o}ta, Bal{\'a}zs and Ahlawat, Paramvir and Cs{\'a}nyi, G{\'a}bor and Simoncelli, Michele},
  journal={arXiv preprint arXiv:2408.00755},
  year={2024}
}

@article{ase-paper,
  author={Ask Hjorth Larsen and Jens Jørgen Mortensen and Jakob Blomqvist and Ivano E Castelli and Rune Christensen and Marcin
Dułak and Jesper Friis and Michael N Groves and Bjørk Hammer and Cory Hargus and Eric D Hermes and Paul C Jennings and Peter
Bjerre Jensen and James Kermode and John R Kitchin and Esben Leonhard Kolsbjerg and Joseph Kubal and Kristen
Kaasbjerg and Steen Lysgaard and Jón Bergmann Maronsson and Tristan Maxson and Thomas Olsen and Lars Pastewka and Andrew
Peterson and Carsten Rostgaard and Jakob Schiøtz and Ole Schütt and Mikkel Strange and Kristian S Thygesen and Tejs
Vegge and Lasse Vilhelmsen and Michael Walter and Zhenhua Zeng and Karsten W Jacobsen},
  title={The atomic simulation environment—a Python library for working with atoms},
  journal={Journal of Physics: Condensed Matter},
  volume={29},
  number={27},
  pages={273002},
  url={http://stacks.iop.org/0953-8984/29/i=27/a=273002},
  year={2017},
}

@article{phonopy-phono3py-JPSJ,
  author  = {Togo, Atsushi},
  title   = {First-principles Phonon Calculations with Phonopy and Phono3py},
  journal = {J. Phys. Soc. Jpn.},
  volume  = {92},
  number  = {1},
  pages   = {012001},
  year    = {2023},
  doi     = {10.7566/JPSJ.92.012001}
}

@misc{rhodesOrbv3AtomisticSimulation2025,
  title = {Orb-v3: Atomistic Simulation at Scale},
  author = {Rhodes, Benjamin and Vandenhaute, Sander and {\v S}imkus, Vaidotas and Gin, James and Godwin, Jonathan and Duignan, Tim and Neumann, Mark},
  year = {2025},
  month = apr,
  number = {arXiv:2504.06231},
  eprint = {2504.06231},
  primaryclass = {cond-mat},
  publisher = {arXiv},
  doi = {10.48550/arXiv.2504.06231},
  archiveprefix = {arXiv}
}

@article{loewUniversalMachineLearning2025,
  title = {Universal Machine Learning Interatomic Potentials Are Ready for Phonons},
  author = {Loew, Antoine and Sun, Dewen and Wang, Hai-Chen and Botti, Silvana and Marques, Miguel A. L.},
  year = {2025},
  month = jun,
  journal = {npj Computational Materials},
  volume = {11},
  number = {1},
  pages = {178},
  publisher = {Nature Publishing Group},
  issn = {2057-3960},
  doi = {10.1038/s41524-025-01650-1},
  copyright = {2025 The Author(s)}
}

@article{dengSystematicSofteningUniversal2025,
  title = {Systematic Softening in Universal Machine Learning Interatomic Potentials},
  author = {Deng, Bowen and Choi, Yunyeong and Zhong, Peichen and Riebesell, Janosh and Anand, Shashwat and Li, Zhuohan and Jun, KyuJung and Persson, Kristin A. and Ceder, Gerbrand},
  year = {2025},
  month = jan,
  journal = {npj Computational Materials},
  volume = {11},
  number = {1},
  pages = {9},
  publisher = {Nature Publishing Group},
  issn = {2057-3960},
  doi = {10.1038/s41524-024-01500-6},
  copyright = {2025 The Author(s)}
}

@inproceedings{fu2025learning,
  title = 	 {Learning Smooth and Expressive Interatomic Potentials for Physical Property Prediction},
  author =       {Fu, Xiang and Wood, Brandon M and Barroso-Luque, Luis and Levine, Daniel S. and Gao, Meng and Dzamba, Misko and Zitnick, C. Lawrence},
  booktitle = 	 {Proceedings of the 42nd International Conference on Machine Learning},
  pages = 	 {17875--17893},
  year = 	 {2025},
  editor = 	 {Singh, Aarti and Fazel, Maryam and Hsu, Daniel and Lacoste-Julien, Simon and Berkenkamp, Felix and Maharaj, Tegan and Wagstaff, Kiri and Zhu, Jerry},
  volume = 	 {267},
  series = 	 {Proceedings of Machine Learning Research},
  month = 	 {13--19 Jul},
  publisher =    {PMLR},
  url = 	 {https://proceedings.mlr.press/v267/fu25h.html},
}

@article{bochkarevGraphAtomicCluster2024,
  title = {Graph {{Atomic Cluster Expansion}} for {{Semilocal Interactions}} beyond {{Equivariant Message Passing}}},
  author = {Bochkarev, Anton and Lysogorskiy, Yury and Drautz, Ralf},
  year = {2024},
  month = jun,
  journal = {Physical Review X},
  volume = {14},
  number = {2},
  pages = {021036},
  publisher = {American Physical Society},
  doi = {10.1103/PhysRevX.14.021036}
}

@article{riebesellFrameworkEvaluateMachine2025,
  title = {A Framework to Evaluate Machine Learning Crystal Stability Predictions},
  author = {Riebesell, Janosh and Goodall, Rhys E. A. and Benner, Philipp and Chiang, Yuan and Deng, Bowen and Ceder, Gerbrand and Asta, Mark and Lee, Alpha A. and Jain, Anubhav and Persson, Kristin A.},
  year = {2025},
  month = jun,
  journal = {Nature Machine Intelligence},
  volume = {7},
  number = {6},
  pages = {836--847},
  publisher = {Nature Publishing Group},
  issn = {2522-5839},
  doi = {10.1038/s42256-025-01055-1},
  copyright = {2025 The Author(s)}
}

@article{schmidt_machine_2023,
  title={Machine-learning-assisted determination of the global zero-temperature phase diagram of materials},
  author={Schmidt, Jonathan and Hoffmann, Noah and Wang, Hai-Chen and Borlido, Pedro and Carri{\c{c}}o, Pedro JMA and Cerqueira, Tiago FT and Botti, Silvana and Marques, Miguel AL},
  journal={Advanced Materials},
  volume={35},
  number={22},
  pages={2210788},
  year={2023},
  publisher={Wiley Online Library}
}

@article{mehl_aflow_2017,
  title={The AFLOW library of crystallographic prototypes: part 1},
  author={Mehl, Michael J and Hicks, David and Toher, Cormac and Levy, Ohad and Hanson, Robert M and Hart, Gus and Curtarolo, Stefano},
  journal={Computational Materials Science},
  volume={136},
  pages={S1--S828},
  year={2017},
  publisher={Elsevier}
}

@article{merchant_scaling_2023,
  title={Scaling deep learning for materials discovery},
  author={Merchant, Amil and Batzner, Simon and Schoenholz, Samuel S and Aykol, Muratahan and Cheon, Gowoon and Cubuk, Ekin Dogus},
  journal={Nature},
  volume={624},
  number={7990},
  pages={80--85},
  year={2023},
  publisher={Nature Publishing Group UK London}
}

@article{deng_chgnet_2023,
  title={CHGNet as a pretrained universal neural network potential for charge-informed atomistic modelling},
  author={Deng, Bowen and Zhong, Peichen and Jun, KyuJung and Riebesell, Janosh and Han, Kevin and Bartel, Christopher J and Ceder, Gerbrand},
  journal={Nature Machine Intelligence},
  volume={5},
  number={9},
  pages={1031--1041},
  year={2023},
  publisher={Nature Publishing Group UK London}
}

@article{chen_universal_2022,
  title={A universal graph deep learning interatomic potential for the periodic table},
  author={Chen, Chi and Ong, Shyue Ping},
  journal={Nature Computational Science},
  volume={2},
  number={11},
  pages={718--728},
  year={2022},
  publisher={Nature Publishing Group US New York}
}

@article{bartel_critical_2020,
  title={A critical examination of compound stability predictions from machine-learned formation energies},
  author={Bartel, Christopher J and Trewartha, Amalie and Wang, Qi and Dunn, Alexander and Jain, Anubhav and Ceder, Gerbrand},
  journal={npj Computational Materials},
  volume={6},
  number={1},
  pages={97},
  year={2020},
  publisher={Nature Publishing Group UK London}
}

@article{bartel_review_2022,
  title={Review of computational approaches to predict the thermodynamic stability of inorganic solids},
  author={Bartel, Christopher J},
  journal={Journal of Materials Science},
  volume={57},
  number={23},
  pages={10475--10498},
  year={2022},
  publisher={Springer}
}

@article{wang_predicting_2021,
  title={Predicting stable crystalline compounds using chemical similarity},
  author={Wang, Hai-Chen and Botti, Silvana and Marques, Miguel AL},
  journal={npj Computational Materials},
  volume={7},
  number={1},
  pages={12},
  year={2021},
  publisher={Nature Publishing Group UK London}
}

@article{yang2024mattersim,
  title={Mattersim: A deep learning atomistic model across elements, temperatures and pressures},
  author={Yang, Han and Hu, Chenxi and Zhou, Yichi and Liu, Xixian and Shi, Yu and Li, Jielan and Li, Guanzhi and Chen, Zekun and Chen, Shuizhou and Zeni, Claudio and others},
  journal={arXiv preprint arXiv:2405.04967},
  year={2024}
}

@article{batatia2023foundation,
    author = {Batatia, Ilyes and Benner, Philipp and Chiang, Yuan and Elena, Alin M. and Kovács, Dávid P. and Riebesell, Janosh and Advincula, Xavier R. and Asta, Mark and Avaylon, Matthew and Baldwin, William J. and Berger, Fabian and Bernstein, Noam and Bhowmik, Arghya and Bigi, Filippo and Blau, Samuel M. and Cărare, Vlad and Ceriotti, Michele and Chong, Sanggyu and Darby, James P. and De, Sandip and Della Pia, Flaviano and Deringer, Volker L. and Elijošius, Rokas and El-Machachi, Zakariya and Fako, Edvin and Falcioni, Fabio and Ferrari, Andrea C. and Gardner, John L. A. and Gawkowski, Mikołaj J. and Genreith-Schriever, Annalena and George, Janine and Goodall, Rhys E. A. and Grandel, Jonas and Grey, Clare P. and Grigorev, Petr and Han, Shuang and Handley, Will and Heenen, Hendrik H. and Hermansson, Kersti and Ho, Cheuk Hin and Hofmann, Stephan and Holm, Christian and Jaafar, Jad and Jakob, Konstantin S. and Jung, Hyunwook and Kapil, Venkat and Kaplan, Aaron D. and Karimitari, Nima and Kermode, James R. and Kourtis, Panagiotis and Kroupa, Namu and Kullgren, Jolla and Kuner, Matthew C. and Kuryla, Domantas and Liepuoniute, Guoda and Lin, Chen and Margraf, Johannes T. and Magdău, Ioan-Bogdan and Michaelides, Angelos and Moore, J. Harry and Naik, Aakash A. and Niblett, Samuel P. and Norwood, Sam Walton and O’Neill, Niamh and Ortner, Christoph and Persson, Kristin A. and Reuter, Karsten and Rosen, Andrew S. and Rosset, Louise A. M. and Schaaf, Lars L. and Schran, Christoph and Shi, Benjamin X. and Sivonxay, Eric and Stenczel, Tamás K. and Sutton, Christopher and Svahn, Viktor and Swinburne, Thomas D. and Tilly, Jules and van der Oord, Cas and Vargas, Santiago and Varga-Umbrich, Eszter and Vegge, Tejs and Vondrák, Martin and Wang, Yangshuai and Witt, William C. and Wolf, Thomas and Zills, Fabian and Csányi, Gábor},
    title = {A foundation model for atomistic materials chemistry},
    journal = {The Journal of Chemical Physics},
    volume = {163},
    number = {18},
    pages = {184110},
    year = {2025},
    month = {11},
    issn = {0021-9606},
    doi = {10.1063/5.0297006},
    url = {https://doi.org/10.1063/5.0297006},
}

@article{ong_python_2013,
  title={Python Materials Genomics (pymatgen): A robust, open-source python library for materials analysis},
  author={Ong, Shyue Ping and Richards, William Davidson and Jain, Anubhav and Hautier, Geoffroy and Kocher, Michael and Cholia, Shreyas and Gunter, Dan and Chevrier, Vincent L and Persson, Kristin A and Ceder, Gerbrand},
  journal={Computational Materials Science},
  volume={68},
  pages={314--319},
  year={2013},
  publisher={Elsevier}
}

@article{oc20,
author = {Chanussot, Lowik and Das, Abhishek and Goyal, Siddharth and Lavril, Thibaut and Shuaibi, Muhammed and Riviere, Morgane and Tran, Kevin and Heras-Domingo, Javier and Ho, Caleb and Hu, Weihua and Palizhati, Aini and Sriram, Anuroop and Wood, Brandon and Yoon, Junwoong and Parikh, Devi and Zitnick, C. Lawrence and Ulissi, Zachary},
title = {Open Catalyst 2020 (OC20) Dataset and Community Challenges},
journal = {ACS Catalysis},
volume = {11},
number = {10},
pages = {6059-6072},
year = {2021},
doi = {10.1021/acscatal.0c04525},
URL = { 
        https://doi.org/10.1021/acscatal.0c04525
},
eprint = { 
        https://doi.org/10.1021/acscatal.0c04525
}
}

@article{tran_open_2023,
  title={The Open Catalyst 2022 (OC22) dataset and challenges for oxide electrocatalysts},
  author={Tran, Richard and Lan, Janice and Shuaibi, Muhammed and Wood, Brandon M and Goyal, Siddharth and Das, Abhishek and Heras-Domingo, Javier and Kolluru, Adeesh and Rizvi, Ammar and Shoghi, Nima and others},
  journal={ACS Catalysis},
  volume={13},
  number={5},
  pages={3066--3084},
  year={2023},
  publisher={ACS Publications}
}

@article{sriram_open_2024,
author = {Sriram, Anuroop and Choi, Sihoon and Yu, Xiaohan and Brabson, Logan M. and Das, Abhishek and Ulissi, Zachary and Uyttendaele, Matt and Medford, Andrew J. and Sholl, David S.},
title = {The Open DAC 2023 Dataset and Challenges for Sorbent Discovery in Direct Air Capture},
journal = {ACS Central Science},
volume = {10},
number = {5},
pages = {923-941},
year = {2024},
doi = {10.1021/acscentsci.3c01629},
URL = {
        https://doi.org/10.1021/acscentsci.3c01629
},
eprint = {
        https://doi.org/10.1021/acscentsci.3c01629
}
}

@article{saal_materials_2013,
	title = {Materials Design and Discovery with High-Throughput Density Functional Theory: The Open Quantum Materials Database ({OQMD})},
	volume = {65},
	issn = {1543-1851},
	url = {https://doi.org/10.1007/s11837-013-0755-4},
	doi = {10.1007/s11837-013-0755-4},
	shorttitle = {Materials Design and Discovery with High-Throughput Density Functional Theory},
	pages = {1501--1509},
	number = {11},
	journal = {{JOM}},
    year = {2013},
	shortjournal = {{JOM}},
	author = {Saal, James E. and Kirklin, Scott and Aykol, Muratahan and Meredig, Bryce and Wolverton, C.},
	urldate = {2024-10-09},
	date = {2013-11-01},
	langid = {english},
}

@article{curtarolo_aflow_2012,
	title = {{AFLOW}: An automatic framework for high-throughput materials discovery},
	volume = {58},
	issn = {0927-0256},
	url = {https://www.sciencedirect.com/science/article/pii/S0927025612000717},
	doi = {10.1016/j.commatsci.2012.02.005},
	shorttitle = {{AFLOW}},
	pages = {218--226},
	journaltitle = {Computational Materials Science},
	journal = {Computational Materials Science},
	author = {Curtarolo, Stefano and Setyawan, Wahyu and Hart, Gus L. W. and Jahnatek, Michal and Chepulskii, Roman V. and Taylor, Richard H. and Wang, Shidong and Xue, Junkai and Yang, Kesong and Levy, Ohad and Mehl, Michael J. and Stokes, Harold T. and Demchenko, Denis O. and Morgan, Dane},
	urldate = {2024-10-09},
	date = {2012-06-01},
        year = {2012}
}

@article{jain2013commentary,
    author = {Jain, Anubhav and Ong, Shyue Ping and Hautier, Geoffroy and Chen, Wei and Richards, William Davidson and Dacek, Stephen and Cholia, Shreyas and Gunter, Dan and Skinner, David and Ceder, Gerbrand and Persson, Kristin A.},
    title = {Commentary: The Materials Project: A materials genome approach to accelerating materials innovation},
    journal = {APL Materials},
    volume = {1},
    number = {1},
    pages = {011002},
    year = {2013},
    month = {07},
    issn = {2166-532X},
    doi = {10.1063/1.4812323},
    url = {https://doi.org/10.1063/1.4812323},
}

@article{wang_framework_2021,
	title = {A framework for quantifying uncertainty in {DFT} energy corrections},
	volume = {11},
	rights = {2021 The Author(s)},
	issn = {2045-2322},
	url = {https://www.nature.com/articles/s41598-021-94550-5},
	doi = {10.1038/s41598-021-94550-5},
	pages = {15496},
	number = {1},
	journal = {Scientific Reports},
	shortjournal = {Sci Rep},
	author = {Wang, Amanda and Kingsbury, Ryan and {McDermott}, Matthew and Horton, Matthew and Jain, Anubhav and Ong, Shyue Ping and Dwaraknath, Shyam and Persson, Kristin A.},
        year = {2021},
	note = {Publisher: Nature Publishing Group},
}

@article{sun_strongly_2015,
  title={Strongly constrained and appropriately normed semilocal density functional},
  author={Sun, Jianwei and Ruzsinszky, Adrienn and Perdew, John P},
  journal={Physical Review Letters},
  volume={115},
  number={3},
  pages={036402},
  year={2015},
  publisher={APS}
}

@article{furness_accurate_2020,
  title={Accurate and numerically efficient r2SCAN meta-generalized gradient approximation},
  author={Furness, James W and Kaplan, Aaron D and Ning, Jinliang and Perdew, John P and Sun, Jianwei},
  journal={The Journal of Physical Chemistry Letters},
  volume={11},
  number={19},
  pages={8208--8215},
  year={2020},
  publisher={ACS Publications}
}

@article{kresse_ultrasoft_1999,
	title = {From ultrasoft pseudopotentials to the projector augmented-wave method},
	volume = {59},
	url = {https://link.aps.org/doi/10.1103/PhysRevB.59.1758},
	doi = {10.1103/PhysRevB.59.1758},
	pages = {1758--1775},
	number = {3},
	journal = {Physical Review B},
	shortjournal = {Phys. Rev. B},
    year = {1999},
	author = {Kresse, G. and Joubert, D.},
	urldate = {2024-10-09},
	date = {1999-01-15},
	note = {Publisher: American Physical Society},
}

@article{perdew_restoring_2008,
  title={Restoring the density-gradient expansion for exchange in solids and surfaces},
  author={Perdew, John P and Ruzsinszky, Adrienn and Csonka, G{\'a}bor I and Vydrov, Oleg A and Scuseria, Gustavo E and Constantin, Lucian A and Zhou, Xiaolan and Burke, Kieron},
  journal={Physical review letters},
  volume={100},
  number={13},
  pages={136406},
  year={2008},
  publisher={APS}
}

@article{goodall2022rapid,
  title={Rapid discovery of stable materials by coordinate-free coarse graining},
  author={Goodall, Rhys EA and Parackal, Abhijith S and Faber, Felix A and Armiento, Rickard and Lee, Alpha A},
  journal={Science Advances},
  volume={8},
  number={30},
  pages={eabn4117},
  year={2022},
  publisher={American Association for the Advancement of Science}
}

@inproceedings{Batatia2022mace,
author = {Batatia, Ilyes and Kovacs, David P and Simm, Gregor and Ortner, Christoph and Csanyi, Gabor},
 booktitle = {Advances in Neural Information Processing Systems},
 editor = {S. Koyejo and S. Mohamed and A. Agarwal and D. Belgrave and K. Cho and A. Oh},
 pages = {11423--11436},
 publisher = {Curran Associates, Inc.},
 title = {MACE: Higher Order Equivariant Message Passing Neural Networks for Fast and Accurate Force Fields},
 url = {https://proceedings.neurips.cc/paper_files/paper/2022/file/4a36c3c51af11ed9f34615b81edb5bbc-Paper-Conference.pdf},
 volume = {35},
 year = {2022}
}

@article{kingsbury2022performance,
  title={Performance comparison of r2-SCAN and SCAN metaGGA density functionals for solid materials via an automated, high-throughput computational workflow},
  author={Kingsbury, Ryan and Gupta, Ayush S and Bartel, Christopher J and Munro, Jason M and Dwaraknath, Shyam and Horton, Matthew and Persson, Kristin A},
  journal={Physical Review Materials},
  volume={6},
  number={1},
  pages={013801},
  year={2022},
  publisher={APS}
}

@article{Kresse1994,
author = {Kresse, Georg and Hafner, J{\"{u}}rgen},
doi = {10.1103/PhysRevB.49.14251},
isbn = {0163-1829 (Print)$\backslash$n0163-1829 (Linking)},
issn = {0163-1829},
journal = {Physical Review B},
number = {20},
pages = {14251--14269},
pmid = {10010505},
title = {{Ab initio molecular-dynamics simulation of the liquid-metal–amorphous-semiconductor transition in germanium}},
volume = {49},
year = {1994}
}

@article{Kresse1996a,
author = {Kresse, Georg and Furthm{\"{u}}ller, J{\"{u}}rgen},
doi = {10.1103/PhysRevB.54.11169},
isbn = {1098-0121},
issn = {0163-1829},
journal = {Physical Review B},
number = {16},
pages = {11169--11186},
pmid = {9984901},
title = {{Efficient iterative schemes for ab initio total-energy calculations using a plane-wave basis set}},
volume = {54},
year = {1996}
}

@article{perdew1996generalized,
  title={Generalized gradient approximation made simple},
  author={Perdew, John P and Burke, Kieron and Ernzerhof, Matthias},
  journal={Physical Review Letters},
  volume={77},
  number={18},
  pages={3865},
  year={1996},
  publisher={APS}
}

@article{zitnick2020introduction,
  title={An introduction to electrocatalyst design using machine learning for renewable energy storage},
  author={Zitnick, C Lawrence and Chanussot, Lowik and Das, Abhishek and Goyal, Siddharth and Heras-Domingo, Javier and Ho, Caleb and Hu, Weihua and Lavril, Thibaut and Palizhati, Aini and Riviere, Morgane and others},
  journal={arXiv preprint arXiv:2010.09435},
  year={2020}
}

@article{singh2019robust,
  title={Robust and synthesizable photocatalysts for CO2 reduction: a data-driven materials discovery},
  author={Singh, Arunima K and Montoya, Joseph H and Gregoire, John M and Persson, Kristin A},
  journal={Nature Communications},
  volume={10},
  number={1},
  pages={443},
  year={2019},
  publisher={Nature Publishing Group UK London}
}

@article{choudhary2022recent,
  title={Recent advances and applications of deep learning methods in materials science},
  author={Choudhary, Kamal and DeCost, Brian and Chen, Chi and Jain, Anubhav and Tavazza, Francesca and Cohn, Ryan and Park, Cheol Woo and Choudhary, Alok and Agrawal, Ankit and Billinge, Simon JL and others},
  journal={npj Computational Materials},
  volume={8},
  number={1},
  pages={59},
  year={2022},
  publisher={Nature Publishing Group UK London}
}

@article{schwalbe2021differentiable,
  title={Differentiable sampling of molecular geometries with uncertainty-based adversarial attacks},
  author={Schwalbe-Koda, Daniel and Tan, Aik Rui and G{\'o}mez-Bombarelli, Rafael},
  journal={Nature Communications},
  volume={12},
  number={1},
  pages={5104},
  year={2021},
  publisher={Nature Publishing Group UK London}
}

@article{li2023exploiting,
  title={Exploiting redundancy in large materials datasets for efficient machine learning with less data},
  author={Li, Kangming and Persaud, Daniel and Choudhary, Kamal and DeCost, Brian and Greenwood, Michael and Hattrick-Simpers, Jason},
  journal={Nature Communications},
  volume={14},
  number={1},
  pages={7283},
  year={2023},
  publisher={Nature Publishing Group UK London}
}

@article{park2024scalable,
    author = {Park, Yutack and Kim, Jaesun and Hwang, Seungwoo and Han, Seungwu},
    title = {Scalable Parallel Algorithm for Graph Neural Network Interatomic Potentials in Molecular Dynamics Simulations},
    journal = {Journal of Chemical Theory and Computation},
    volume = {20},
    number = {11},
    pages = {4857-4868},
    year = {2024},
    doi = {10.1021/acs.jctc.4c00190},
    note ={PMID: 38813770},
    URL = {   
            https://doi.org/10.1021/acs.jctc.4c00190 
    },
}

@article{sun2016thermodynamic,
  title={The thermodynamic scale of inorganic crystalline metastability},
  author={Sun, Wenhao and Dacek, Stephen T and Ong, Shyue Ping and Hautier, Geoffroy and Jain, Anubhav and Richards, William D and Gamst, Anthony C and Persson, Kristin A and Ceder, Gerbrand},
  journal={Science Advances},
  volume={2},
  number={11},
  pages={e1600225},
  year={2016},
  publisher={American Association for the Advancement of Science}
}

@article{MP-release,
author={The Materials Project},
title={Materials Project Database Versions},
journal={\url{https://docs.materialsproject.org/changes/database-versions}},
year={2024}}

@inproceedings{liao2024equiformerv2,
    title={EquiformerV2: Improved Equivariant Transformer for Scaling to Higher-Degree Representations},
    author={Yi-Lun Liao and Brandon M Wood and Abhishek Das and Tess Smidt},
    booktitle={The Twelfth International Conference on Learning Representations},
    editor={NA},
    year={2024},
    url={https://openreview.net/forum?id=mCOBKZmrzD}
}

@article{liao2024generalizing,
    title={Generalizing Denoising to Non-Equilibrium Structures Improves Equivariant Force Fields},
    author={Yi-Lun Liao and Tess Smidt and Muhammed Shuaibi and Abhishek Das},
    journal={Transactions on Machine Learning Research},
    issn={2835-8856},
    year={2024},
    url={https://openreview.net/forum?id=whGzYUbIWA},
    note={}
}

@misc{pymatgen-issue-2968,
  author = {Materials Project},
  title = {Issue \#2968: Add support for new crystal structure prediction method},
  howpublished = {\url{https://github.com/materialsproject/pymatgen/issues/2968}},
  year = {2023}
}

@misc{pymatgen-issue-3016,
  author = {Materials Project},
  title = {Issue \#3016: Add support for new crystal structure prediction method},
  howpublished = {\url{https://github.com/materialsproject/pymatgen/issues/3016}},
  year = {2023}
}

@misc{riebesell_pymatviz_2022,
  title = {Pymatviz: visualization toolkit for materials informatics},
  author = {Riebesell, Janosh and Yang, Haoyu and Goodall, Rhys and Baird, Sterling G.},
  date = {2022-10-01},
  year = {2022},
  doi = {10.5281/zenodo.7486816},
  url = {https://github.com/janosh/pymatviz},
  note = {10.5281/zenodo.7486816 - https://github.com/janosh/pymatviz},
  version = {0.12.0}
}

@article{horton2023crystal,
  title={Crystal Toolkit: A Web App Framework to Improve Usability and Accessibility of Materials Science Research Algorithms},
  author={Horton, Matthew and Shen, Jimmy-Xuan and Burns, Jordan and Cohen, Orion and Chabbey, Fran{\c{c}}ois and Ganose, Alex M and Guha, Rishabh and Huck, Patrick and Li, Hamming Howard and McDermott, Matthew and others},
  journal={arXiv preprint arXiv:2302.06147},
  year={2023}
}

@article{Jain2013,
author = {Jain, A. and Ong, S. P. and Hautier, G. and Chen, W. and Richards, W. D. and Dacek, S. and Cholia, S. and Gunter, D. and Skinner, D. and Ceder, G. and Persson, K. A.},
doi = {10.1063/1.4812323},
issn = {2166532X},
journal = {APL Materials},
number = {1},
pages = {011002},
title = {{The Materials Project: A materials genome approach to accelerating materials innovation}},
url = {http://link.aip.org/link/AMPADS/v1/i1/p011002/s1\&Agg=doi},
volume = {1},
year = {2013}
}

@article{nequip_2022,
	author = {Batzner, Simon and Musaelian, Albert and Sun, Lixin and Geiger, Mario and Mailoa, Jonathan P. and Kornbluth, Mordechai and Molinari, Nicola and Smidt, Tess E. and Kozinsky, Boris},
	journal = {Nature Communications},
	number = {1},
	pages = {2453},
	title = {E(3)-equivariant graph neural networks for data-efficient and accurate interatomic potentials},
	volume = {13},
	year = {2022}}

@article{choudhary_alignn,
	author = {Choudhary, Kamal and DeCost, Brian},
	journal = {npj Computational Materials},
	number = {1},
	pages = {185},
	title = {Atomistic Line Graph Neural Network for improved materials property predictions},
	volume = {7},
	year = {2021}}

@article{ceder_1998,
	author = {Ceder, G. and Chiang, Y. -M. and Sadoway, D. R. and Aydinol, M. K. and Jang, Y. -I. and Huang, B.},
	journal = {Nature},
	number = {6677},
	pages = {694--696},
	title = {Identification of cathode materials for lithium batteries guided by first-principles calculations},
	volume = {392},
	year = {1998}}

@article{biomat_2015,
	author = {Mark W. Tibbitt and Christopher B. Rodell and Jason A. Burdick and Kristi S. Anseth},
	journal = {Proceedings of the National Academy of Sciences},
	number = {47},
	pages = {14444-14451},
	title = {Progress in material design for biomedical applications},
	volume = {112},
	year = {2015}}

@techreport{carbtree_comp_mat,
  author       = {Crabtree, George and Glotzer, Sharon and McCurdy, Bill and Roberto, Jim},
  title        = {Computational Materials Science and Chemistry:  Accelerating Discovery and Innovation through Simulation-Based Engineering and Science},
  url          = {https://www.osti.gov/biblio/1294275},
  place        = {United States},
  year         = {2010},
  month        = {07},
}

@article{pogue_2023,
	author = {Pogue, Elizabeth A. and New, Alexander and McElroy, Kyle and Le, Nam Q. and Pekala, Michael J. and McCue, Ian and Gienger, Eddie and Domenico, Janna and Hedrick, Elizabeth and McQueen, Tyrel M. and Wilfong, Brandon and Piatko, Christine D. and Ratto, Christopher R. and Lennon, Andrew and Chung, Christine and Montalbano, Timothy and Bassen, Gregory and Stiles, Christopher D.},
	journal = {npj Computational Materials},
	number = {1},
	pages = {181},
	title = {Closed-loop superconducting materials discovery},
	volume = {9},
	year = {2023}}
\clearpage
\newpage
\beginappendix
\counterwithin{figure}{section}
\counterwithin{table}{section}
\section{OMat24 psuedopotential differences} \label{sec:psuedo-diffs}

Differences between OMat24 DFT and Materials Project PBE calculations are largely attributed to different versions of pseudo-potentials and different pseudo-potential symbols for Yb and W. In OMat24 we use version 54 of VASP pseudopotentials, while the Materials Project calculations use an older releases of VASP pseudopotentials\cite{horton2025accelereated}. Table \ref{tab:potcars} lists element symbols and pseudopotential generation dates for all elements with different pseudo potentials.

\begin{table}[h!]
\centering
\begin{tabular}{cll}
\hline
\textbf{Element} & \textbf{MP (version 52)}                & \textbf{OMat (version 54)}            \\ \hline
Cl   & Cl 17Jan2003          & Cl 06Sep2000             \\ \hline
S    & S 17Jan2003           & S 06Sep2000              \\ \hline
Au   & Au 06Sep2000          & Au 04Oct2007             \\ \hline
W    & W\_pv 06Sep2000       & W\_sv 04Sep2015          \\ \hline
Rh   & Rh\_pv 06Sep2000      & Rh\_pv 25Jan2005         \\ \hline
Pt   & Pt 05Jan2001          & Pt 04Feb2005             \\ \hline
Ru   & Ru\_pv 06Sep2000      & Ru\_pv 28Jan2005         \\ \hline
Pd   & Pd 05Jan2001          & Pd 04Jan2005             \\ \hline
P    & P 17Jan2003           & P 06Sep2000              \\ \hline
As   & As 06Sep2000          & As 22Sep2009             \\ \hline
Mo   & Mo\_pv 08Apr2002      & Mo\_pv 04Feb2005         \\ \hline
Ge   & Ge\_d 06Sep2000       & Ge\_d 03Jul2007          \\ \hline
Ag   & Ag 06Sep2000          & Ag 02Apr2005             \\ \hline
Tc   & Tc\_pv 06Sep2000      & Tc\_pv 04Feb2005         \\ \hline
Co   & Co 06Sep2000          & Co 02Aug2007             \\ \hline
Fe   & Fe\_pv 06Sep2000      & Fe\_pv 02Aug2007         \\ \hline
Ga   & Ga\_d 06Sep2000       & Ga\_d 06Jul2010          \\ \hline
Cr   & Cr\_pv 07Sep2000      & Cr\_pv 02Aug2007         \\ \hline
Mn   & Mn\_pv 07Sep2000      & Mn\_pv 02Aug2007         \\ \hline
Zr   & Zr\_sv 07Sep2000      & Zr\_sv 04Jan2005         \\ \hline
Mg   & Mg\_pv 06Sep2000      & Mg\_pv 13Apr2007         \\ \hline
Y    & Y\_sv 06Sep2000       & Y\_sv 25May2007          \\ \hline
Gd   & Gd 08Apr2002          & Gd 23Dec2003             \\ \hline
Eu   & Eu 08Apr2002          & Eu 23Dec2003             \\ \hline
Ce   & Ce 28Sep2000          & Ce 23Dec2003             \\ \hline
Yb   & Yb\_2 06Sep2000       & Yb\_3 08Jul2013          \\ \hline
Li   & Li\_sv 23Jan2001      & Li\_sv 10Sep2004         \\ \hline
Na   & Na\_pv 05Jan2001      & Na\_pv 19Sep2006         \\ \hline
\end{tabular}
\caption{\textbf{OMat24 and MP POTCAR file differences.} POTCAR file symbols and titel dates for all elements that with different POTCARs in Materials Project and OMat24 PBE calculations. \label{tab:potcars}}
\end{table}

\section{Dataset statistics}

\begin{figure}[h]
    \centering
    \includegraphics[width=\textwidth]{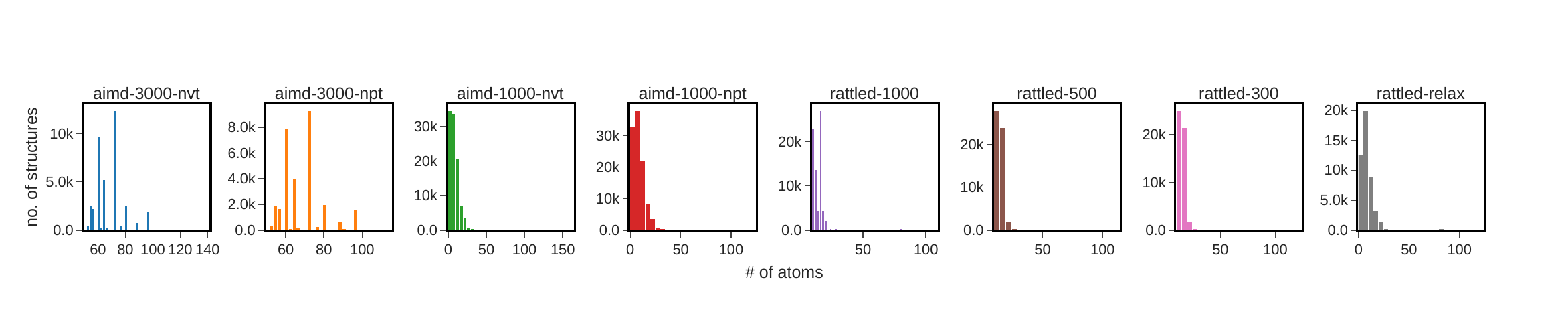}
    \caption{\textbf{Distribution of atoms per structure across OMat24.} Histogram of number of atoms per structure per sub-dataset in OMat-24 dataset.}
    \label{fig:omat-sub-natoms}
\end{figure}

\begin{figure}[ht]
    \centering
    \includegraphics[width=\textwidth]{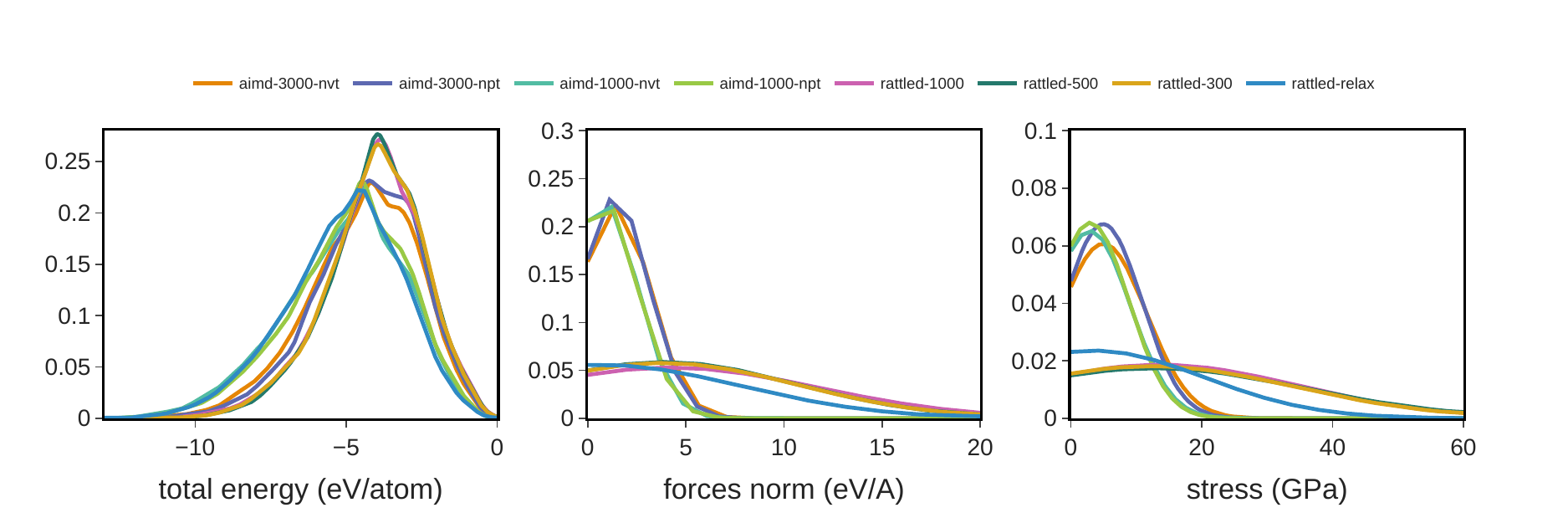}
    \caption{\textbf{Density distributions across OMat24 sub-datasets.} Density distrubtions showing energy, forces norm and maximum absolute stress densities for all sub-datasets in OMat-24.}
    \label{fig:omat-sub-stats}
\end{figure}

Figure \ref{fig:omat-sub-natoms} shows histograms for the number of atoms per structure in the sub-datasets that make up the OMat24 dataset. Similarly Figure \ref{fig:omat-sub-stats} shows the distributions of energy, forces and stress in the different sub-datasets. The number of atoms per system is shifted to largers sizes, since these structures were generated by tiling the unit cell of relaxed Alexandria structures. The rest of the distributions are concentrated at smaller sizes, as a direct result of the distribution in the Alexandria dataset shown in Figure \ref{fig:omat-overview}b in the main text.

From Figure \ref{fig:omat-sub-stats} we can see that the AIMD portion of the dataset has a much narrower force and stress distribution compared to that of structures generated from rattled relaxations and rattled Boltzmann sampling. We also observe a shift to lower energies for structures generated using AIMD.

\begin{figure}
    \centering
    \includegraphics[width=\textwidth]{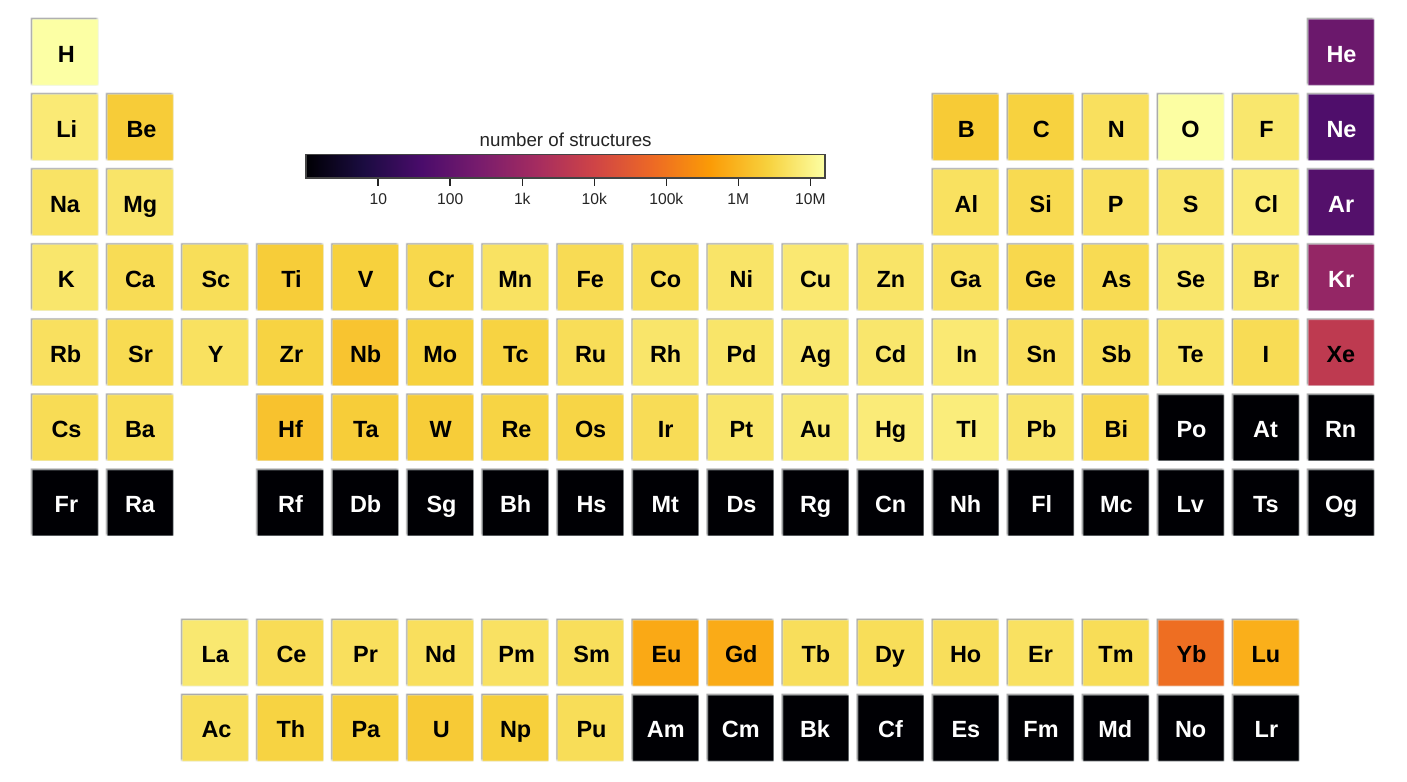}
    \caption{\textbf{Element distribution in Alexandria PBE dataset.}\cite{schmidt_machine_2023}.}
    \label{fig:alex-elements}
\end{figure}

\begin{figure}
    \centering
    \includegraphics[width=\textwidth]{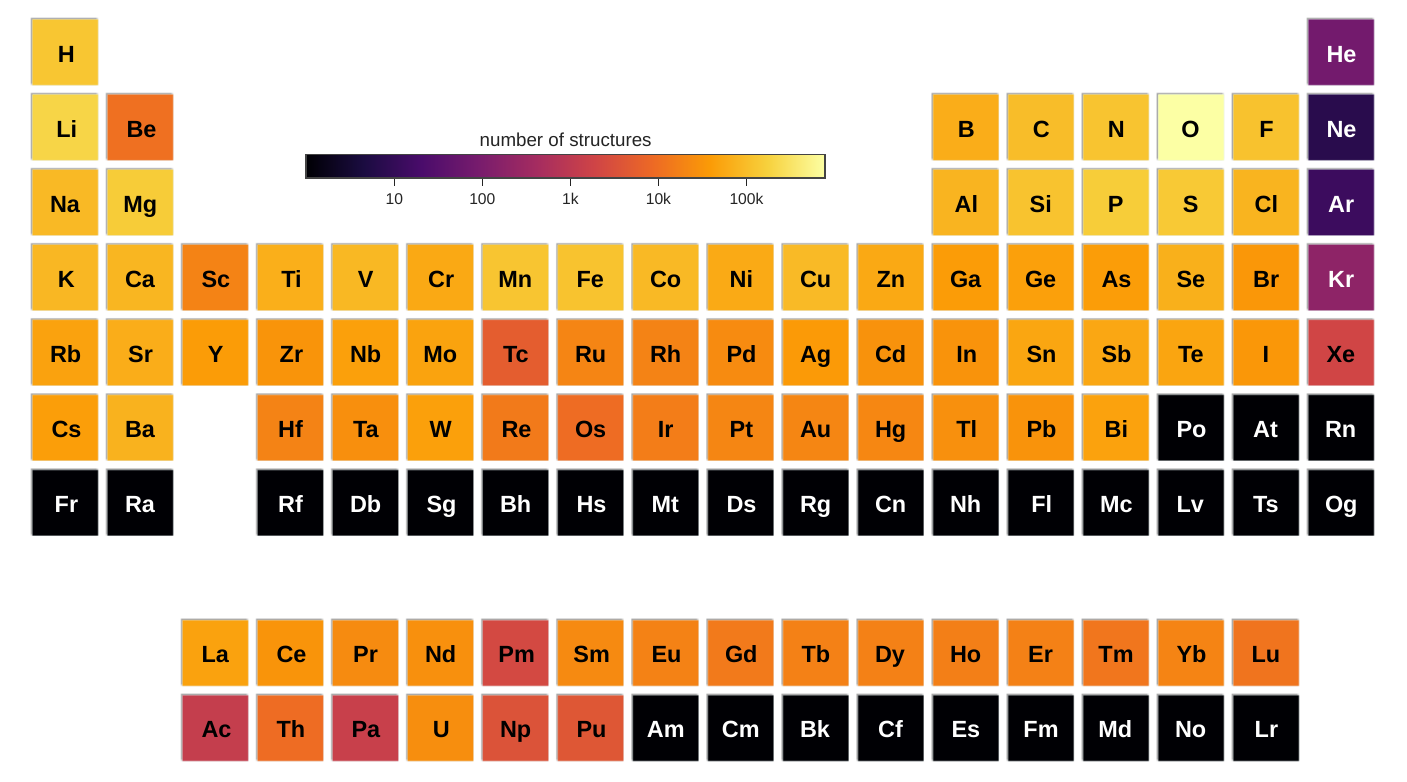}
    \caption{\textbf{Element distribution in MPtrj dataset.} \cite{deng_chgnet_2023}.}
    \label{fig:mptrj-elements}
\end{figure}

Figures \ref{fig:alex-elements} and \ref{fig:mptrj-elements} show per element histograms of the number of occurences in each dataset. The histograms are created by randomly sampling 1\% and 10\% of each dataset respectively, as a result elements that appear very rarely may have been missed. By comparing Figures \ref{fig:alex-elements} and \ref{fig:mptrj-elements} with the corresponding histogram for the OMat24 dataset in Figure \ref{fig:omat-overview}c we observe that the distributions between OMat24 and Alex are very similar and more uniformly distributed that that of MPtrj.

\begin{figure}
    \centering
    \includegraphics[width=\textwidth]{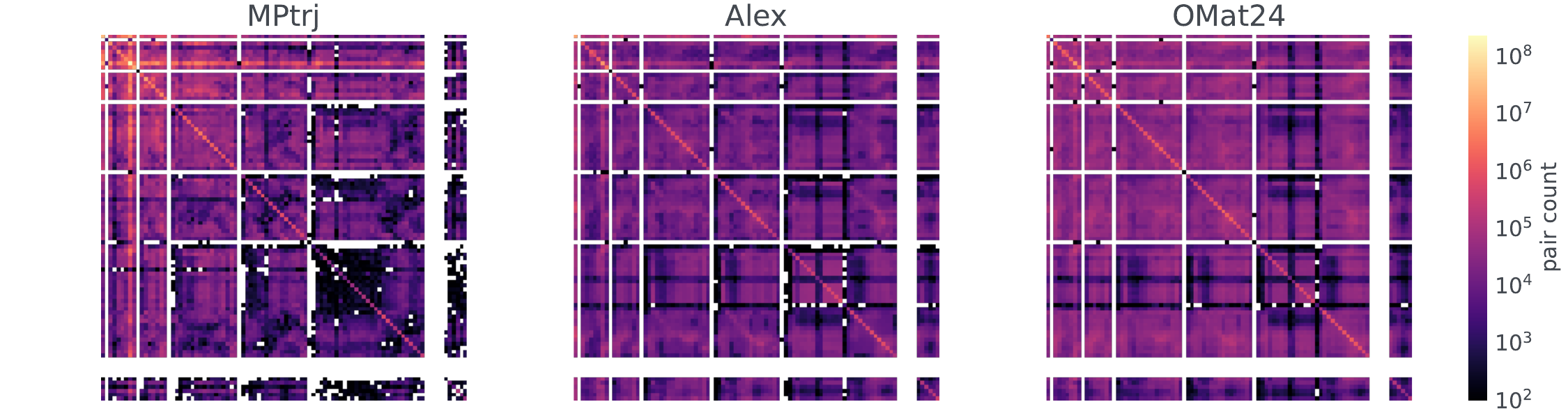}
    \caption{\textbf{Number of distinct pair interactions for atoms within 3.5\AA.}}
    \label{fig:pair-counts}
\end{figure}

The composition and structural diversity of the OMat24 dataset are further demonstrated by examining the variety of local atomic neighborhoods. Figure \ref{fig:pair-counts} displays the total number of distinct element pairs within a 3.5\AA~radius for the OMat24, Alexandria, and MPtrj datasets. The pair counts for Ale4xandria and OMat24 are several orders of magnitude greater than those for MPtrj, reflecting the relative sizes of the datasets. Furthermore, the OMat24 dataset exhibits more uniform and comprehensive coverage of near-neighbor pairs compared to both Alexandria and MPtrj.

\subsection{Drift forces in AIMD data}

\begin{figure}
    \centering
    \includegraphics[width=0.5\textwidth]{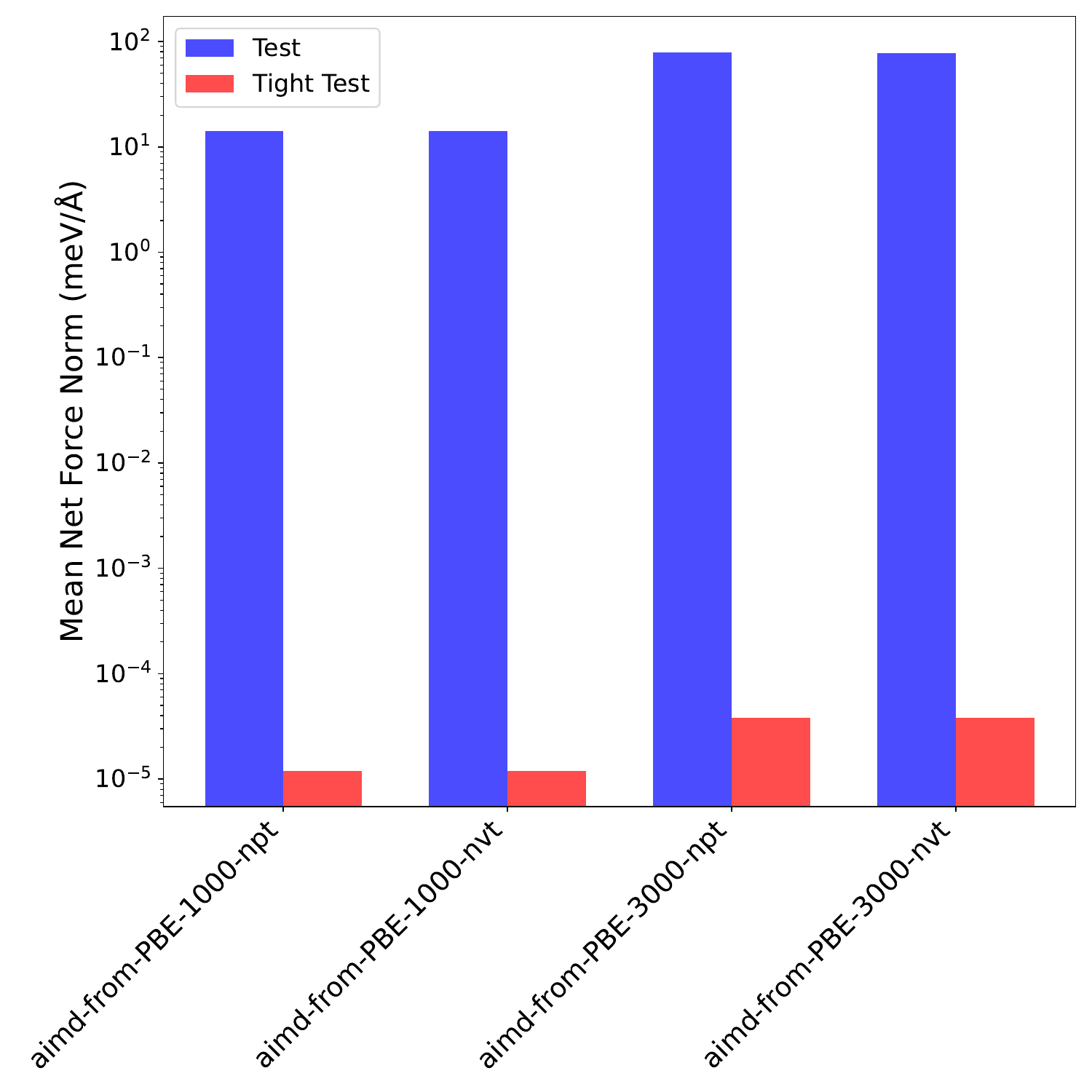}
    \caption{\textbf{Mean drift forces in AIMD.} Mean drift (net forces) per atom for AIMD 1M test subsplits calculated with different energy convergence thresholds. (Test) OMat24 threshold, (Tight Test) stricter $1\times 10^{-6}$ threshold.}
    \label{fig:drift-forces}
\end{figure}

Nonnegligible drift forces—also referred to as nonzero net forces—can arise in Ab Initio Molecular Dynamics (AIMD) calculations. VASP routines are written to correct the drift and ensure dynamics consistent with the ensemble being used. However, the forces printed in output files of AIMD calculations include the drift terms. A recent study found that AIMD data in the OMat24 dataset included several structures with substantial drift forces (up to 1 eV/Å) \cite{kuryla2025accuratedftforces}. Figure \ref{fig:drift-forces} presents the mean net force per atom for AIMD subsplits in the OMat24-1M test set. Consistent with the independent study \cite{kuryla2025accuratedftforces}, net forces can reach up to 1 meV/\AA in AIMD calculations when using OMat24 settings. We also recomputed the data as single point calculations of the AIMD frames and with a much stricter electronic convergence (EDIFF=$1\times 10^{-6}$), and observe that the net forces are no longer present.

Following similar analyses \cite{sahoo2025opencatalyst2025oc25}, we find that machine-learned interatomic potentials (MLIPs) trained on OMat24 as-is predict tightly converged DFT forces with accuracy comparable to, or even better than, predictions for less tightly converged calculations. Table \ref{tab:force-convergence-model-val} presents the test MAE for eSEN and eqV2-S models on both the regularly converged and tightly converged test sets, showing that errors are effectively equivalent. This result aligns with previous experiments and suggests that MLIP models can learn the true forces without being affected by this level of numerical noise from drift forces. We do not advocate for future data generation with looser convergence thresholds. Rather, future efforts should aim to balance computational efficiency with fully converged calculations considering the effects of artifacts such as drift-forces\cite{sahoo2025opencatalyst2025oc25, kaplanFoundationalPotentialEnergy2025a, kunerMPALOER2SCANDataset2025}.

\begin{table*}[!htp]\centering 
\caption{\textbf{Model prediction performance vs DFT convergence.} Model prediction MAE over a subsample of the test dataset and the same data points computed tighter convergence settings.}\label{tab:force-convergence-model-val}
\begin{tabular}{ccccc}
\hline
Model & \multicolumn{2}{c}{Energy MAE (meV/atom)} & \multicolumn{2}{c}{Force MAE (meV/\AA)} \\
& Regular & Tight & Regular & Tight \\
\hline
eSEN &9.0 &6.9 &40.1 &37.9 \\
eqV2 S &10.0 &8.1 &41.9 &39.9 \\
\hline
\end{tabular}
\end{table*}

\clearpage

\section{OMat24 elemental references and formation energy corrections} \label{sec:omat-refs}

In order to calculate formation energies and energies above hull involving DFT calculations using GGA and GGA with Hubbard U corrections, the Materials Project has developed a mixing compatibility correction \cite{wang_framework_2021}. In addition, an energy adjustment to certain anion species is added to account for electron localization errors \cite{wang_framework_2021}. These compatiblity corrections are obtained by way of a least squares fit minimizing the error between predicted formation energies and a set of experimental formation energies with reported uncertainties below 100 meV/atom \cite{wang_framework_2021}.

We have followed the same procedure and computed a compatiblity correction set for OMat24 DFT settings. We use the same 222 compounds predicted to be within 100 meV of the MP convex hull that are used in fitting the MP2020 Compatiblity corrections. The +U mixing and anion correction values per elememnt/species are listed in Tables \ref{tab:ggau-corrections} and \ref{tab:anion-corrections} respectivelty.

\begin{table*}[!htp]\centering 
\caption{\textbf{GGA/GGA+U transition metal mixing corrections.} Fitted GGA+U mixing energy corrections and their associated uncertainties for transition metal oxides and fluorides (in eV/atom). The corrections were fit following the MP2020 Compatibility scheme \cite{wang_framework_2021}.\label{tab:ggau-corrections}}
\begin{tabular}{ccccc}
\hline
Species & OMat24 Correction & OMat24 Uncertainty & MP Correction & MP Uncertainty \\
\hline
V  & -1.813 & 0.0064 & -1.7   & 0.0064 \\
Cr & -2.037 & 0.0108 & -1.999 & 0.0108 \\
Mn & -1.701 & 0.0053 & -1.668 & 0.0053 \\
Fe & -2.428 & 0.0101 & -2.256 & 0.0101 \\
Co & -2.151 & 0.006  & -1.638 & 0.006  \\
Ni & -2.58  & 0.0107 & -2.541 & 0.0107 \\
W  & -4.445 & 0.0253 & -4.438 & 0.0253 \\
Mo & -2.972 & 0.0089 & -3.202 & 0.0089 \\
\hline
\end{tabular}
\end{table*}

\begin{table*}[!htp]\centering 
\caption{\textbf{GGA/GGA+U anion mixing corrections.} Fitted Composition-based energy corrections and uncertainties for various anions (in eV/atom) . The corrections were fit following the MP2020 Compatibility scheme \cite{wang_framework_2021}\label{tab:anion-corrections}}
\begin{tabular}{ccccc}
\hline
Anion/Element & OMat24 Correction & OMat24 Uncertainty & MP Correction & MP Uncertainty \\
\hline
oxide      & -0.657 & 0.002  & -0.687 & 0.002  \\
peroxide   & -0.433 & 0.0172 & -0.465 & 0.0172 \\
superoxide & -0.152 & 0.0075 & -0.161 & 0.0075 \\
S          & -0.487 & 0.0093 & -0.503 & 0.0093 \\
F          & -0.436 & 0.0026 & -0.462 & 0.0026 \\
Cl         & -0.6   & 0.0018 & -0.614 & 0.0018 \\
Br         & -0.318 & 0.0026 & -0.534 & 0.0026 \\
I          & -0.194 & 0.0055 & -0.379 & 0.0055 \\
N          & -0.303 & 0.0093 & -0.361 & 0.0093 \\
Se         & -0.474 & 0.0341 & -0.472 & 0.0341 \\
Si         & 0.028  & 0.0165 & 0.071  & 0.0165 \\
Sb         & -0.194 & 0.0089 & -0.192 & 0.0089 \\
Te         & -0.418 & 0.0262 & -0.422 & 0.0262 \\
H          & -0.173 & 0.0013 & -0.179 & 0.0013 \\
ozonide    & 0      & 0      & 0      & 0      \\
\hline
\end{tabular}
\end{table*}

\begin{figure}[h]
    \centering
    \includegraphics[width=0.5\textwidth]{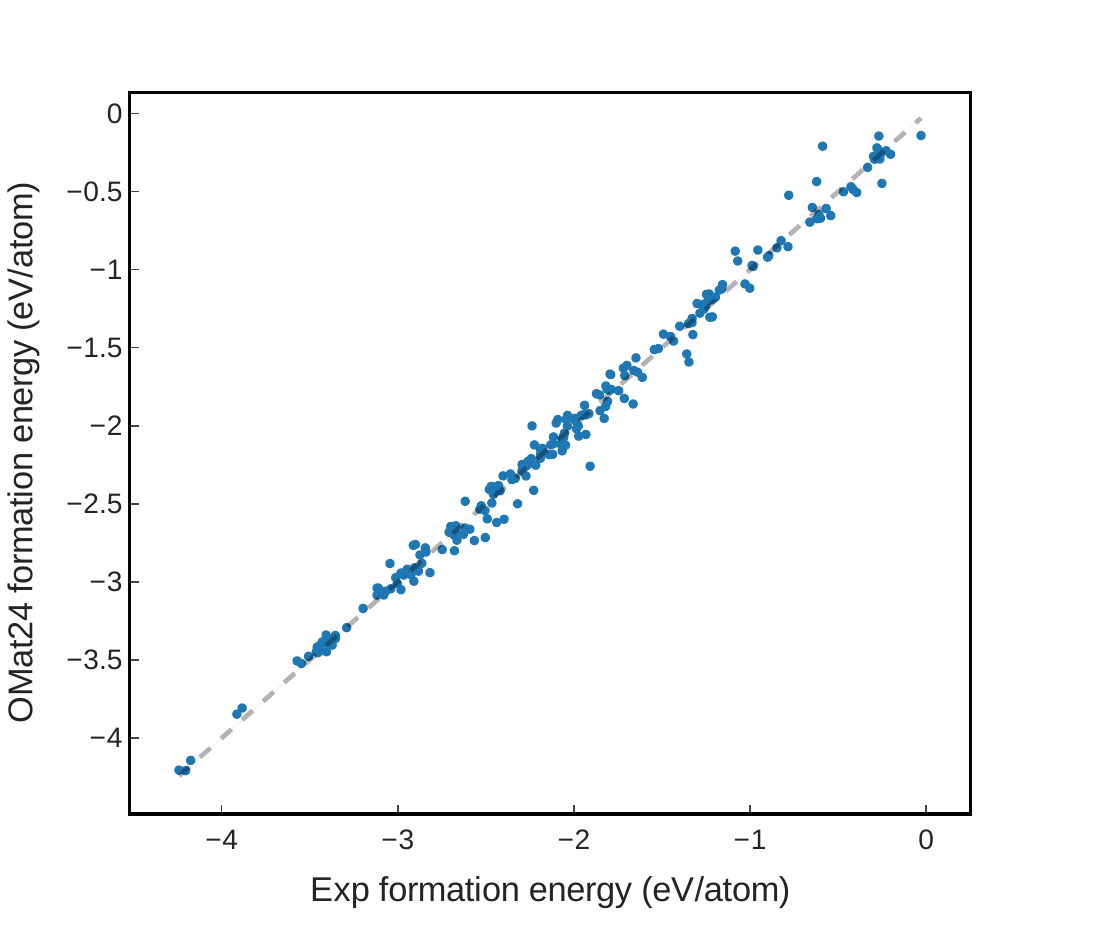}
    \caption{\textbf{OMat24 DFT vs experimental formation energy parity.} Formation energy calculated with OMat24 DFT settings vs experimental formation energy.}
    \label{fig:omat-exp-parity}
\end{figure}

\begin{figure}[ht]
    \centering
    \includegraphics[width=\textwidth]{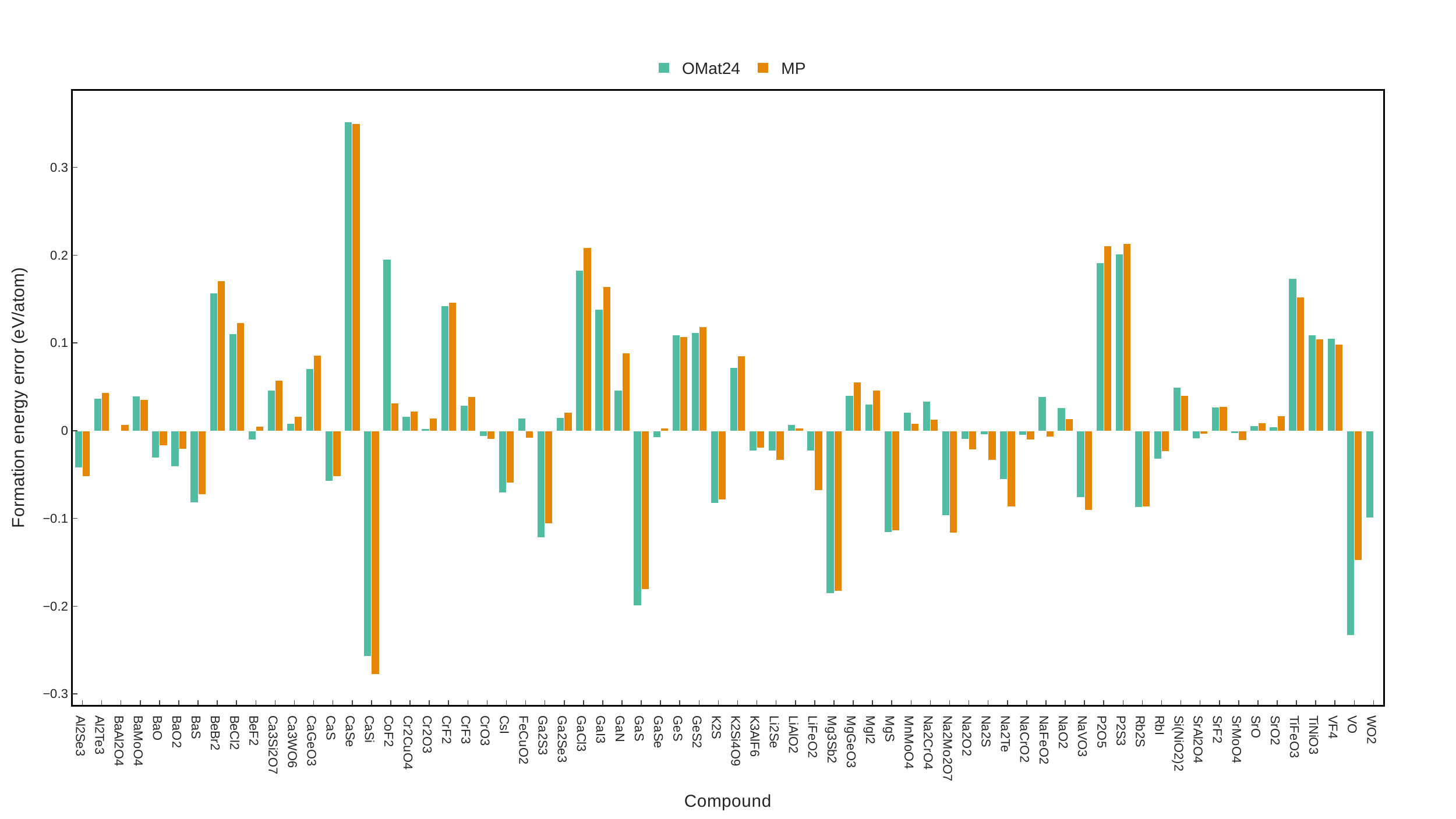}
    \caption{\textbf{OMat24 DFT ormation energy error with respect to experiment.} Formation energy error between OMat24 DFT settings and experimentally reported values for selected compounds with the highest reported measurement uncertainty (>10meV/atom)}
    \label{fig:e-form-exp-error}
\end{figure}

The parity plot of the DFT predicted formation energies and experimental formation energies is shown in Figure \ref{fig:omat-exp-parity}. The resulting mean absolute error between the DFT predicted and experimental formation energies for the 222 compounds is 55 meV/atom. Figure \ref{fig:e-form-exp-error} shows the difference in predicted and experiemntal formation energy for compounds with the largest experimental formation energy uncertainties (> 10 meV/atom). 

\begin{figure}
    \centering
    \includegraphics[width=\textwidth]{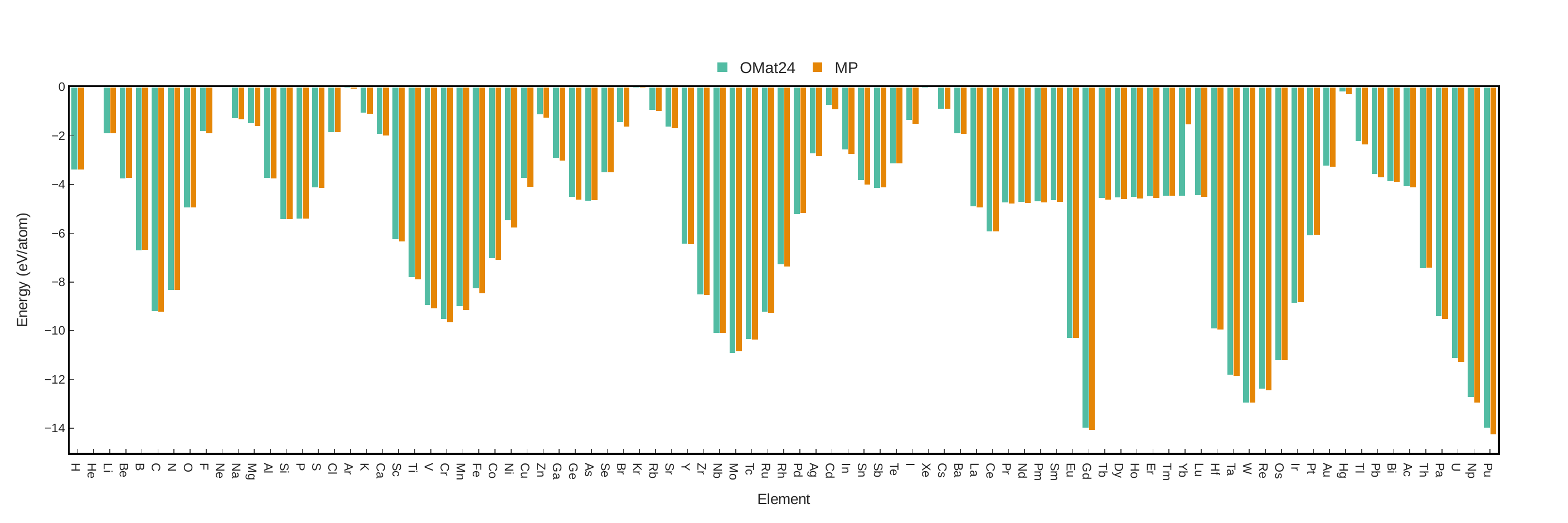}
    \caption{\textbf{Comparison of elemental reference energies between OMat24 and Materials Project DFT settings.}}
    \label{fig:element-refs}
\end{figure}

In order compute formation energies using OMat24 DFT settings or models trained on OMat24 only, we have also computed the energy of elemental reference compounds. Figure \ref{fig:element-refs} shows the energy of elemental reference compounds computed with the OMat24 DFT settings compared to the PBE calculations in MP. We reiterate that the OMat24 references should be used when computing formation energies with OMat24 trained models. Similary, we recommend that the OMat24 compatiblity corrections be used when computing formation energies involving compounds for which a +U correction was or would have been used in the DFT calculation accordingly. Furthermore, we remind readers that when using MP2020 style compatiblity corrections, the predicted formation energies are meant to represent values at room temperature (the temperature of experimental formation energies used to fit the corrections). We refer readers to the original publication \cite{wang_framework_2021} detailing the correction scheme to get a full understanding of the details and nuances involved in these calculations.

\clearpage

\section{Dataset Learning Curves}

\begin{figure}[h!]
    \centering
    \includegraphics[width=\linewidth]{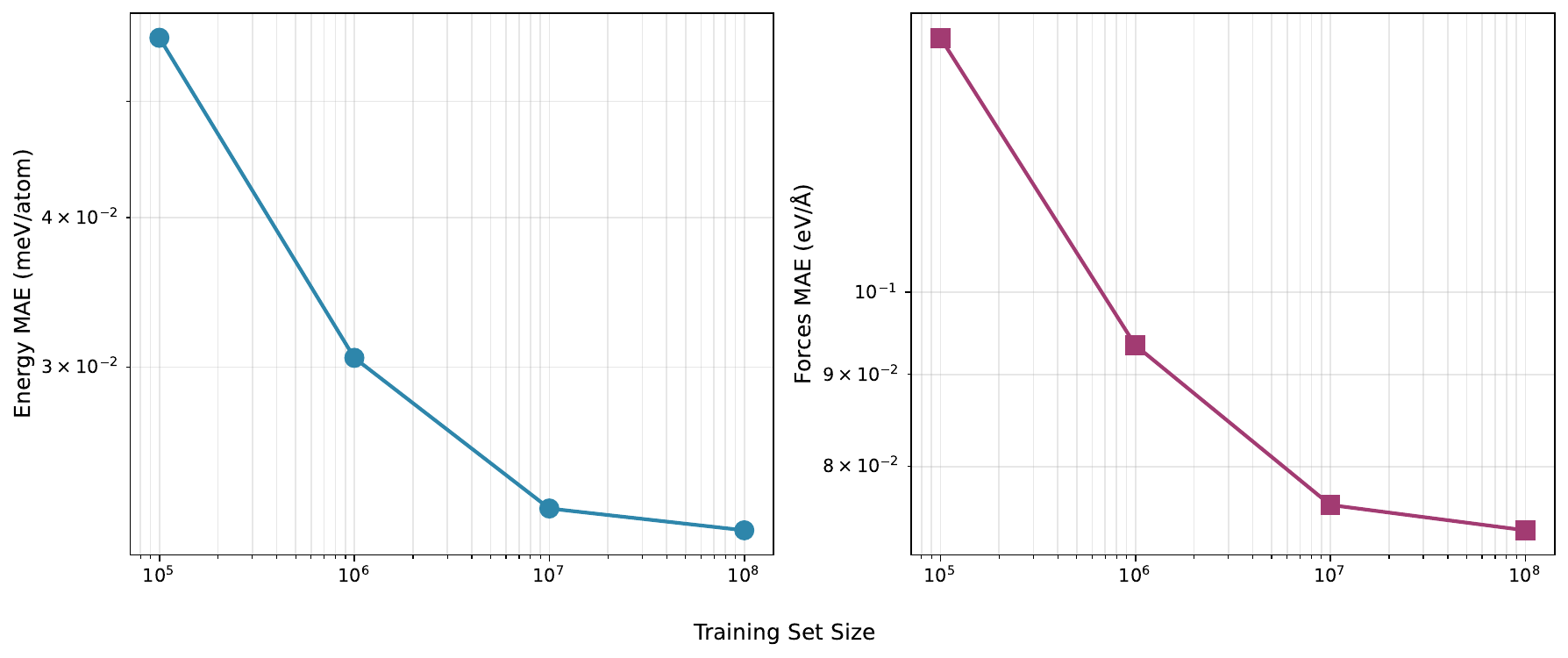}
    \caption{\textbf{Learning curves for energy and force predictions of the eSEN model on OMat24 subsamples.} Energy mean absolute error (MAE) and forces MAE as a function of number of structures in training set.}
    \label{fig:learning-curves}
\end{figure}

Figure \ref{fig:learning-curves} presents the energy and force mean absolute error (MAE) learning curves for OMat24 training subsamples of increasing size using the eSEN architecture. Consistent with general machine learning principles, we observe that prediction errors decrease as the dataset size increases, provided the model has sufficient capacity. Notably, for the eSEN architecture in these experiments, the error plateaus at approximately 100 million samples, which matches the total size of the OMat24 training set.

\section{Systematic softening improvements across model architectures}

\begin{figure}[h!]
    \centering
    \includegraphics[width=\linewidth]{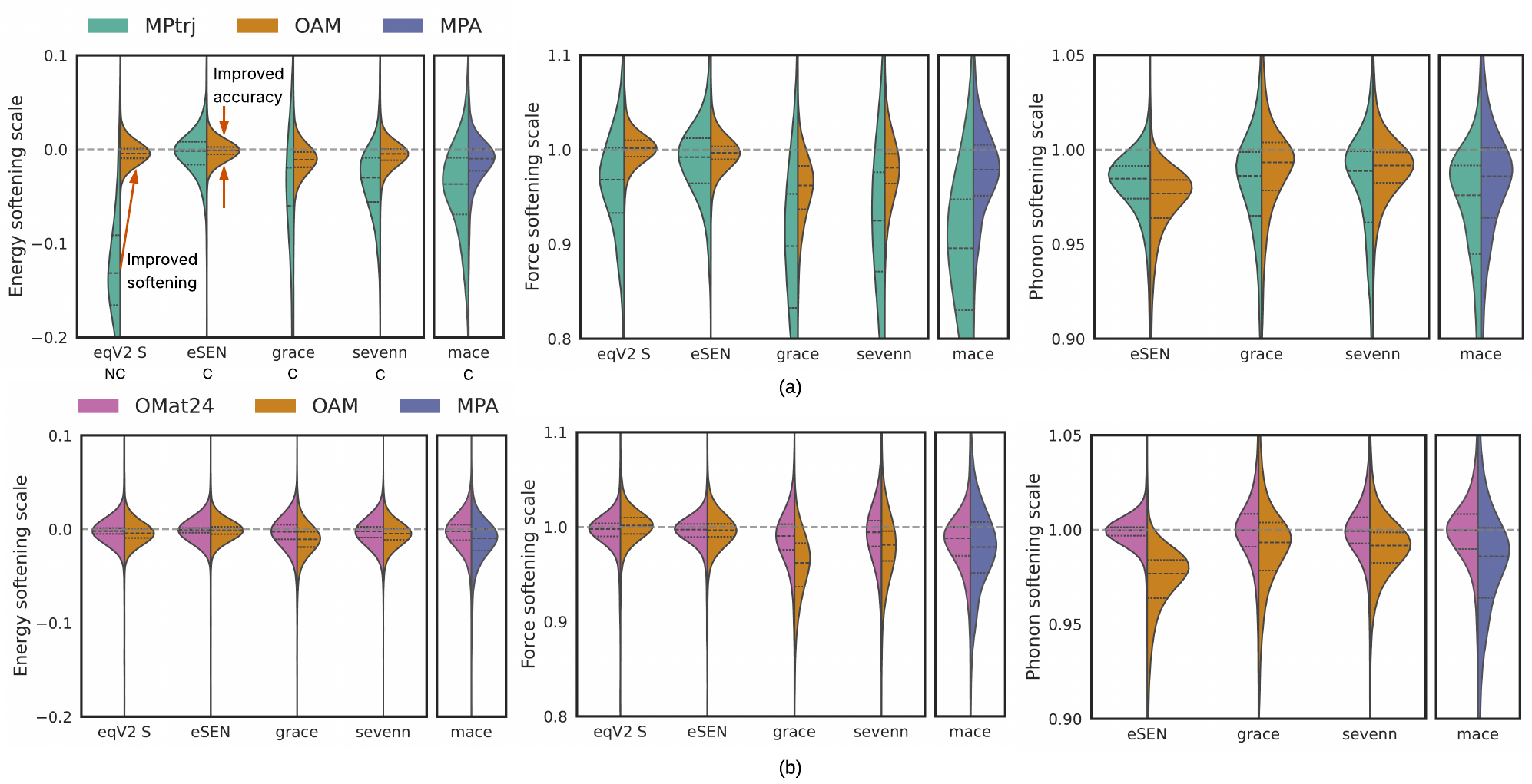}
    \caption{\textbf{Systematic softening improvements across additional model architectures.} Zeroth order (energy), first order (forces) and second order (phonon) systematic softening distribution shifts between (a) MPTrj, OAM (OMat24 + MP/sAlex finetuning) and MPA (MP/sAlex) trained models and (b) OMat24, OAM, and MPA models. Violin plots show the distribution of the energy and force softening scale of model predictions on high energy WBM set \cite{dengSystematicSofteningUniversal2025}, and phonon softening of conservative model predictions on the recalculated MDR phonon dataset \cite{loewUniversalMachineLearning2025}. Models are labeled (C) \emph{conservative} if forces are predicted as energy gradients, and (NC) \emph{not conservative} otherwise.}
    \label{fig:softening-all}
\end{figure}

Figure \ref{fig:softening-all} presents violin plots illustrating the various orders of softening, as introduced in the main text, for two additional architectures: GRACE and Sevenn. Improvements across all degrees of softening are consistently observed in five distinct architectures, indicating that the effect is largely architecture-independent and primarily attributable to the diversity of the OMat24 dataset.

\section{Effects of denoising augmentation versus dataset diversity}

We investigate whether denoising augmentation can produce accurate models with smaller datasets and further enhance accuracy when combined with OMat24.

To assess the impact of denoising augmentation and establish a performance baseline, we utilize the denoising nonequilibrium structures (DeNS) protocol \cite{liao2024generalizing}. We compare models trained exclusively on the MPtraj dataset, both with and without DeNS, which are classified as \emph{compliant} in the Matbench Discovery benchmark \cite{riebesellFrameworkEvaluateMachine2025}. This allows a structured comparison of architectures and the effects of DeNS on a fixed dataset. Subsequently, we also evaluate the impact of training on the OMat24 dataset, again with and without DeNS.

\begin{table*}[!htp]\centering 
\caption{\textbf{Matbench-Discovery benchmark results of compliant models.} benchmark results of compliant models trained only on MPtrj with results on the unique prototype split. Mean absolute error (MAE) and Root mean squared error (RMSE) are in units of eV/atom.\label{tab:mbd-compliant}. Top third party models from Matbench-Discovery snapshot before May 1, 2025.}
\scalebox{0.825}{
\begin{tabular}{l|ccccc|ccc}\toprule
Model &eSEN &eqV2-L-DeNS &eqV2-M-DeNS &eqV2-S-DeNS &eqV2-S &MatRIS v0.5.0 &ORB v2 &SevenNet \\\midrule
F1 $\uparrow$ &\textbf{0.831} &0.823 &0.818 &0.815 &0.77 &0.809 &0.765 &0.724 \\
DAF $\uparrow$ &\textbf{5.260} &5.184 &5.109 &5.042 &4.64 &5.049 &4.702 &4.252 \\
Precision $\uparrow$ &\textbf{0.804} &0.792 &0.781 &0.771 &0.709 &0.772 &0.719 &0.65 \\
Recall $\uparrow$ &\textbf{0.861} &0.856 &0.858 &0.864 &0.841 &0.850 &0.817 &0.818 \\
Accuracy $\uparrow$ &\textbf{0.946} &0.944 &0.942 &0.941 &0.926 &0.938 &0.922 &0.904 \\ \midrule
TPR $\uparrow$ &0.861 &0.856 &0.858 &\textbf{0.864} &0.841 &0.850 &0.817 &0.818 \\
FPR $\downarrow$ &\textbf{0.038} &0.041 &0.044 &0.047 &0.063 &0.46 &0.059 &0.081 \\
TNR $\uparrow$ &\textbf{0.962} &0.959 &0.956 &0.953 &0.937 &0.954 &0.941 &0.919 \\
FNR $\downarrow$ &0.139 &0.144 &0.142 &\textbf{0.136} &0.159 &0.150 &0.183 &0.182 \\ \midrule
MAE $\downarrow$ &\textbf{33} &35 &35 &36 &42 &37 &45 &48  \\
RMSE $\downarrow$ &\textbf{78} &82 &82 &85 &87 &82 &91 &92 \\
R2 $\uparrow$ &\textbf{0.822} &0.802 &0.803 &0.788 &0.778 &0.803 &0.756 &0.75 \\
\bottomrule
\end{tabular}}
\end{table*}

Table \ref{tab:mbd-compliant} presents the results of the best compliant models on the \emph{discovery} task of the Matbench Discovery benchmark. The eSEN and equiformerV2 models with DeNS outperform their non-DeNS counterparts and all other third-party models in this task. For the larger equiformerV2 model, DeNS regularization facilitates effective training even with the smaller MPtraj dataset. The energy MAE remains consistent (35-36 meV/atom) across different equiformerV2 sizes, indicating the practical utility of the smaller model. These findings suggest that dataset augmentation techniques like DeNS can significantly enhance performance when training solely on the MPtraj dataset.

\begin{figure*}[h]
    \centering
    \includegraphics[width=\textwidth]{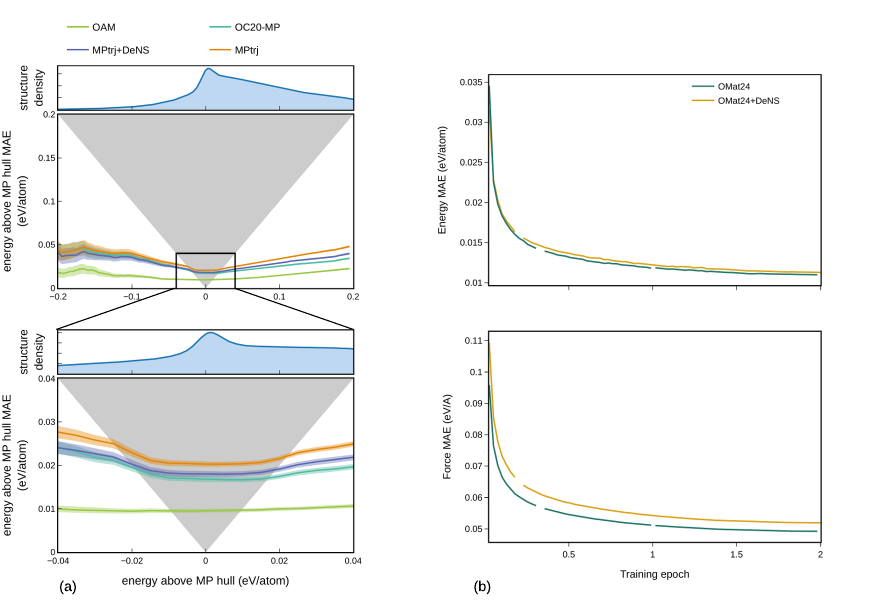}
    \caption{\textbf{Impact of denoising augmentation and dataset diversity on model accuracy and validation errors.} (a) Rolling mean absolute error of predicted energy above hull for equiformerV2 small models (31M parameters) using a window of 40 meV/atom. (b) energy and force mean absolute error validation curves during equiformerV2 small models training on OMat24 with DeNS and without it.} 
    \label{fig:dens-v-data}
\end{figure*}

Figure \ref{fig:dens-v-data} (a) depicts the rolling mean absolute error of energy above hull predictions using a 40 meV/atom window for small equiformerV2 models trained on various datasets. Non-zero errors within the gray triangle indicate that the model error exceeds the true energy above hull, and with probablity leading to incorrect stability classifications.

We observe an overall improvement when training with DeNS on MPtraj compared to using MPtraj alone. More pronounced performance gains are evident in models pre-trained on larger, diverse datasets. Pre-training on OC20, a diverse catalysis dataset, slightly enhances performance over DeNS after fine-tuning on MPtraj for structures near the convex hull. Furthermore, models trained on OMat24, an in-domain dataset, exhibit the most significant performance improvements. Training on OMat24 results in models with a broader window of constant accuracy, indicating better extrapolation to regions with lower dataset density farther away from the convex hull.

Finally, Figure \ref{fig:dens-v-data} (b) presents the energy and force validation mean absolute error during the training of equiformerV2-S models on OMat24, both with and without DeNS. We observe a slight performance penalty when using DeNS with OMat24, indicating that the diversity of the larger OMat24 dataset does not benefit from the regularization or augmentation provided by DeNS.

\section{Model training hyper-parameters and configuration} \label{sec:supp-training}

The models trained using the EquiformerV2 architecture \cite{liao2024equiformerv2} have separate heads for energy, forces and stress prediction. We use a per-atom MAE loss for energy and an $l_2$ norm loss for forces. For stress prediction, we decompose the $3 \times 3$ symmetric stress tensor to a 1-dimensional isotropic component ($L=0$) and a 5-dimensional anisotropic component ($L=2$). We use a prediction head that outputs a scalar and irreps of $L=2$, then use an MAE loss for the isotropic and the anisotropic component separately. At test time, we recover the stress tensor by combining the isotropic and anisotropic components. For models trained with DeNS, an additional head is added to predict the noise added to a perturbed structure. For a DeNS forward pass, we input the noisy structure and predict the unperturbed energy as well as the noise vectors, then compute a per-atom MAE loss for energy and an MSE loss for the noise vectors. We refer interested readers to the original EquiformerV2 and DeNS papers for more details on model architectures\cite{liao2024equiformerv2, liao2024generalizing}. The hyper-parameters for our models are summarized in Table \ref{table:apdx_model_hps}. The hyper-parameters for training are summarized in Table \ref{table:apdx_training_hps}. We use the same mixed precision strategy as the original EquiformerV2 paper\cite{liao2024equiformerv2} to speed up the training process. Models were trained on 64 NVIDIA A100 GPUs for pre-training and 32 NVIDIA A100 GPUs for fine-tuning with distributed data parallel.

Models trained using the eSEN architecture \cite{fu2025learning} are trained using a two step procedure. First a direct force pretraining stage is carried out, where the model has seperate heads for energy, force and stress prediction. Subsequently, a conservative finetuning stage is carried out. In the conservative finetuning state the direct force and stress heads are removed, and force and stress predictions are obtained from autodifferentiation of the energy predictions. This two step training strategy results in models with higher accuracy and requiring less total training epochs and less total training time compared to training a conservative model from scratch \cite{fu2025learning}. For additional training details, hyperparameter values, we refer the reader to the original eSEN publication \cite{fu2025learning}.

\begin{table}[ht]
\centering
\caption{\textbf{Hyper-parameters for the EquiformerV2 models of different sizes.} All hyper-parameters for a given model size is used for all dataset settings. We denote the dimension of irreps features as $(L_{max}, C)$ where $L_{max}$ is the maximum degree and $C$ is the number of channels.}
\label{table:apdx_model_hps}
\scalebox{0.8}{
\begin{tabular}{l|lll}
\toprule
Hyper-parameters & \multicolumn{1}{c}{eqV2-S} & \multicolumn{1}{c}{eqV2-M} & \multicolumn{1}{c}{eqV2-L} \\
\midrule
Maximum degree $L_{max}$ & $4$ & $6$ & $6$ \\
Maximum order $M_{max}$ & $2$ & $4$ & $3$\\
Number of Transformer blocks & $8$ & $10$ & $20$ \\
Cutoff radius (\AA) & $12$ & $12$ & $12$ \\
Maximum number of neighbors & $20$ & $20$ & $20$ \\
Number of radial bases & $600$ & $600$ & $600$ \\
Dimension of hidden scalar features in radial functions $d_{edge}$ & $(0, 128)$ & $(0, 128)$ & $(0, 128)$ \\
Embedding dimension $d_{embed}$ & $(4, 128)$ & $(6, 128)$ & $(6, 128)$ \\
$f_{ij}^{(L)}$ dimension $d_{attn\_hidden}$ & $(4, 64)$ & $(6, 64)$ & $(6, 64)$ \\
Number of attention heads $h$ & $8$ & $8$ & $8$ \\
$f_{ij}^{(0)}$ dimension $d_{attn\_alpha}$ & $(0, 64)$ & $(0, 64)$ & $(0, 64)$ \\
Value dimension $d_{attn\_value}$ & $(4, 16)$ & $(6, 16)$ & $(6, 16)$\\
Hidden dimension in feed forward networks $d_{ffn}$ & $(4, 128)$ & $(6, 128)$ & $(6, 128)$ \\
Resolution of point samples $R$ & $18$ & $18$ & $18$ \\
\bottomrule
\end{tabular}
}
\end{table}

\begin{table}[ht]
\centering
\caption{\textbf{Hyper-parameters for EquiformerV2 model training for different dataset settings.} All model sizes use the same set of hyper-parameters for a given dataset setting.}
\label{table:apdx_training_hps}
\scalebox{0.7}{
\begin{tabular}{l|llll}
\toprule
Hyper-parameters & \multicolumn{1}{c}{MPtrj training} & \multicolumn{1}{c}{OMat training} & \multicolumn{1}{c}{OAM fine-tuning} \\
\midrule
Optimizer & AdamW & AdamW AdamW \\
Learning rate scheduling & Cosine & Cosine & Cosine \\
Warmup epochs & $0.1$ & $0.01$ & $0.1$ \\
Warmup factor & $0.2$ & $0.2$ & $0.2$ \\
Maximum learning rate & $2 \times 10 ^{-4}$ & $6 \times 10 ^{-4}$ & $2 \times 10^{-4}$ \\
Minimum learning rate factor & $0.01$ & $0.01$ & $0.01$ \\
Batch size & $512$ & $512$ & $256$ \\
Number of epochs & $150$ & $2$ & $8$ \\
Gradient clipping norm threshold & $100$ & $100$ & $100$ \\
Model EMA decay & $0.999$ & $0.999$ & $0.999$ \\
Weight decay & $1 \times 10 ^{-3}$ & $1 \times 10 ^{-3}$ & $1 \times 10 ^{-3}$\\
Dropout rate & $0.1$ & $0.1$ & $0.1$ \\
Stochastic depth & $0.1$ & $0.1$ & $0.1$ \\
Energy loss coefficient & $20$ & $20$ & $20$ \\
Force loss coefficient & $20$ & $20$ & $10$ \\
Stress loss coefficient & $5$ & $5$ $1$ \\
\midrule[0.6pt]
DeNS settings \\
\midrule[0.6pt]
Probability of optimizing DeNS & $0.5$ & $0.25$ & -\\
Standard deviation of Gaussian noise & $0.1$ & $0.1$ & - \\
DeNS loss coefficient & $10$ & $10$ & - \\
\bottomrule
\end{tabular}
}
\end{table}

\begin{table}[ht]
\centering
\caption{\textbf{Hyper-parameters for the eSEN models.} eSEN-30M-OMat was trained for 2 epochs using  direct-force pre-training and 2 epochs of conserved fine-tuning. \textsuperscript{\textdagger}The eSEN-30M-OAM model starts from the eSEN-30M-OMat model, and was finetuned for 1 epoch on a dataset constructed by combining the sAlex training dataset and 8 copies of the MPTrj training dataset \cite{fu2025learning}.}
\label{table:apdx_training_hps_esen}
\scalebox{0.8}{
\begin{tabular}{l|ccccccc}
\toprule
Hyper-parameters & MPTrj training & OMat training & OAM Fine-tuning \\
\midrule
Number of eSEN~layer blocks & $10$ & $10$ & $10$ \\
Maximum degree $L_{\mathrm{max}}$ & $3$ & $3$ & $3$\\
Maximum order $M_{\mathrm{max}}$ & $2$ & $2$ & $2$\\
Number of channels $N_{\mathrm{channel}}$ & $128$ & $128$ & $128$ \\
Radial basis function & Gaussian & Gaussian & Gaussian \\
Number of radial basis functions & $10$ & $64$ & $64$ \\
Cutoff radius (\AA) & $6$ & $6$ & $6$ \\
Batch size & 512 & 512 & 256 \\
Optimizer & AdamW & AdamW & AdamW \\
Learning rate scheduling & Cosine & Cosine & Cosine \\
Warmup epochs & $0.1$ & $0.1$ & $0.1$ \\
Warmup factor & $0.2$ & $0.2$ & $0.2$  \\
Maximum learning rate & $4 \times 10 ^{-4}$ & $4 \times 10 ^{-4}$ & $2 \times 10 ^{-4}$ \\
Number of epochs & $60 + 40$\textsuperscript{*} & $2 + 2$\textsuperscript{*} & $1$\textsuperscript{\textdagger} \\
Gradient clipping norm & $100$ & $100$ & $100$ \\
Model EMA decay& $0.999$ & $0.999$ & $0.999$ \\
Weight decay & $1 \times 10 ^{-3}$ & $1 \times 10 ^{-3}$ & $1 \times 10 ^{-3}$ \\
Energy loss coefficient & $20$ & $20$ & $20$ \\
Force loss coefficient & $20$ & $20$ & $20$\\
Stress loss coefficient & $5$ & $5$ & $5$ \\
\bottomrule
\end{tabular}
}
\end{table}

Table \ref{tab:model-sizes} lists the total number of parameters and inference throughput on an Nvidia A100 GPU of the final trained models.
\begin{table}[!htp]\centering
\caption{\textbf{Total number of model parameters and inference throughput.} Total number of parameters and their inference throughput in the EquiformerV2 models in this work. Throughput evaluated on Nvidia A100 GPUs with batch size 1 and no inference-time optimization with samples from the MPtrj dataset. \label{tab:model-sizes}}
\begin{tabular}{c|c|c}\toprule
 & & Throughput\\ Model & \# of Parameters &Samples / GPU sec. (MPtrj) \\\midrule
eqV2-S &31,207,434  & 9.7 \\
eqV2-M &86,589,068 & 7.1\\
eqV2-L &153,7698,68 & 4.5 \\
eSEN &30,161,153 & 5.5 \\
\bottomrule
\end{tabular}
\end{table}

\clearpage
\subsection{Matbench-Discovery metrics}
Table \ref{tab:mbd-metrics-exp} lists descriptions of metrics by abbreviation as used in the Matbench-Discovery benchmark.

\begin{table}[ht!]
\centering
\caption{\textbf{Matbench-Discovery metrics and descriptions}. For further information refer to the original publication.\cite{riebesellFrameworkEvaluateMachine2025}}
\label{tab:mbd-metrics-exp}
\scalebox{0.8}{
\begin{tabular}{l|p{10cm}}
\toprule
\textbf{Metric} & \textbf{Description} \\
\midrule
F1 & Harmonic mean of precision and recall from below hull classification. \\
DAF & Discovery Average Fidelity; ration of below hull classification precision and prevalence. How much more effective a model is compared to random classification. \\
Precision & Ratio of true positives below hull to the sum of true positives and false positives. \\
Recall & Ratio of true positives to the sum of true positives and false negatives. \\
Accuracy & Proportion of true results among the total number of cases examined. \\
TPR & True Positive Rate. \\
FPR & False Positive Rate. \\
TNR & True Negative Rate. \\
FNR & False Negative Rate. \\
MAE & Mean Absolute Error; average of the absolute differences between predicted and actual energy above hull. \\
RMSE & Root Mean Squared Error; square root of the average of the squared differences between predicted and actual energy above hull. \\
R2 & R-squared measure for regression predictions and actual energy above hull values. \\
\bottomrule
\end{tabular}
}
\end{table}

\section{Model test and validation metrics}

\subsection{OMat24 test metrics}
Table \ref{tab:omat-test} list validation and test metrics for models trained using the OMat24 dataset.

\begin{table*}[!htp]\centering
\caption{\textbf{OMat24 test metrics.} Test mean absolute error metrics of equiformerV2 models trained on OMat24 dataset. Energy errors are in units meV/atom, forces errors are in meV/\r{A} and stress errors are in meV/\r{A}$^3$.\label{tab:omat-test}}
\begin{tabular}{l|p{0.7cm}p{0.7cm}p{0.8cm}|p{0.7cm}p{0.7cm}p{0.8cm}|p{0.7cm}p{0.7cm}p{0.8cm}|p{0.7cm}p{0.7cm}p{0.8cm}}\toprule
&\multicolumn{3}{c}{WBM test} &\multicolumn{3}{c}{ID test} &\multicolumn{3}{c}{OOD composition test} &\multicolumn{3}{c}{OOD element test} \\\cmidrule{2-13}
Model &energy &forces &stress&energy &forces &stress &energy &forces &stress  &energy  &forces &stress \\\midrule
eSEN-30M &15.96 &50.68 &3.69 &11.19 &49.33 &2.44 &11.05 &48.85 &2.53 &9.34 &49.14 &2.16 \\
eqV2-S &15.96 &50.68 &3.69 &11.19 &49.33 &2.44 &11.05 &48.85 &2.53 &9.34 &49.14 &2.16 \\
eqV2-M &14.87 &46.26 &3.61 &10.17 &44.90 &2.35  &9.98 &44.51 &2.45 &8.83 &44.71 &2.06 \\
eqV2-L &14.57 &44.72 &3.57 &9.79 &43.21	&2.32  &9.70 &43.00 &2.42 &8.38 &43.08 &2.04 \\
\bottomrule
\end{tabular}
\end{table*}

\section{MPtrj validation metrics}
Tables \ref{tab:mptrj-scratch-val} and \ref{tab:mptrj-ft-val} list the validation metrics for the modeles trained or fine-tuned solely on MPtrj.

\begin{table*}[!htp]\centering
\caption{\textbf{Validation mean absolute error metrics for models trained on the MPtrj dataset.} Energy errors are in units meV/atom, forces errors are in meV/\r{A} and stress errors are in meV/\r{A}$^3$.\label{tab:mptrj-scratch-val}}
\scalebox{0.8}{
\begin{tabular}{l|rrrrr}\toprule
model &energy $\downarrow$ &forces $\downarrow$ &stress $\downarrow$ &forces cos $\uparrow$\\\midrule
eqV2-S &12.4 &32.22 &1.55 &0.72 \\
eqV2-S-DeNS &11.43 &31.67 &1.44 &0.72 \\
eqV2-M-DeNS &11.17 &31.46 &1.48 &0.73 \\
eqV2-L-DeNS &10.58 &30.48 &1.47 &0.74 \\
eSEN &17.02 &43.96 &1.40 &0.72 \\
\bottomrule
\end{tabular}}
\end{table*}
\begin{table*}[!htp]\centering
\caption{\textbf{Validation metrics for finetuning OMat24 pretrained models on MPtrj.}\label{tab:mptrj-ft-val}}
\scalebox{0.8}{
\begin{tabular}{lrrrrr}\toprule
model &energy (meV/atom) $\downarrow$ &forces (meV/A) $\downarrow$ &stress (meV/A$^3$) $\downarrow$ &forces cos $\uparrow$\\\midrule
eqV2-S-OMat-MP &8.52 &23.86 &1.3 &0.764 \\
eqV2-L-OMat-MP &7.99 &22.63 &1.28 &0.777 \\
eSEN &8.33 &23.33 &1.68 &0.783 \\
\bottomrule
\end{tabular}}
\end{table*}

\end{document}